%% file: main.tex
\documentclass{article}

\PassOptionsToPackage{numbers, compress}{natbib}
\usepackage[preprint]{neurips_2025}

\usepackage[utf8]{inputenc} % allow utf-8 input
\usepackage[T1]{fontenc}    % use 8-bit T1 fonts
\usepackage{hyperref}       % hyperlinks
\usepackage{url}            % simple URL typesetting
\usepackage{booktabs}       % professional-quality tables
\usepackage{amsfonts}       % blackboard math symbols
\usepackage{nicefrac}       % compact symbols for 1/2, etc.
\usepackage{microtype}      % microtypography
\usepackage{xcolor}         % colors

\usepackage{times}
\usepackage{latexsym}
\usepackage{amsmath}
\usepackage{amssymb}
\usepackage{amsfonts}
\usepackage{graphicx}
\usepackage{multirow}
\usepackage{verbatim}
\usepackage[justification=centering]{subfig}
\usepackage{tikz}
\usepackage{hyperref}
\usetikzlibrary{fit,positioning}
\usepackage{comment}
\usepackage{float}
\usepackage{cleveref}
\usepackage{amsthm}
\usepackage[linesnumbered,ruled,vlined]{algorithm2e}
\usepackage{multicol}
\usepackage{xcolor}
\usepackage{mathrsfs}
\usepackage{subcaption}
\usepackage{sidecap,caption}
\usepackage{enumitem}
\usepackage{booktabs, multirow}

\title{\emph{Permissioned LLMs}: Enforcing Access Control in Large
  Language Models}

\author{%
  Bargav Jayaraman\\
  Oracle Labs\\
  \texttt{bargav.jayaraman@oracle.com}\\
  \And
  Virendra J. Marathe\\
  Oracle Labs\\
  \texttt{virendra.marathe@oracle.com}\\
  \And
  Hamid Mozaffari\\
  Oracle Labs\\
  \texttt{hamid.mozaffari@oracle.com}\\
  \And
  William F. Shen\\
  University of Cambridge\\
  \texttt{fs604@cam.ac.uk}
  \And
  Krishnaram Kenthapadi\\
  Oracle Health\\
  \texttt{krishnaram.kenthapadi@oracle.com}\\
}

\begin{document}

\maketitle

\input{cmds}

\input{abstract}

\input{intro}

\input{formalism}

\input{algos}

\input{audit}

\input{eval}

\input{discussion}

\input{ack}

\bibliographystyle{plainnat}
\bibliography{refs}

\appendix
\input{appendix}

%%%%%%%%%%%%%%%%%%%%%%%%%%%%%%%%%%%%%%%%%%%%%%%%%%%%%%%%%%%%

\end{document}

%% file: cmds.tex
\newcommand{\sref}[2]{\hyperref[#2]{#1 \ref*{#2}}}

\newtheorem{theorem}{Theorem}[section]
\newtheorem{lemma}[theorem]{Lemma}

\newtheorem{definition}{Definition}[section]

\renewcommand{\sectionautorefname}{\S}
\renewcommand{\subsectionautorefname}{\S}
\renewcommand{\subsubsectionautorefname}{\S}

\newcommand{\dpsgd}{\emph{DP-SGD}}
\newcommand{\dpfedsgd}{\emph{DP-FedSGD}}
\newcommand{\localdpsgd}{\emph{ItemDP}}
\newcommand{\usercentraldp}{\emph{CentralUserDP}}
\newcommand{\userlocaldpsgd}{\emph{ClientLDP}}
\newcommand{\userlocalsgd}{\emph{User-Local-SGD}}
\newcommand{\userlocaloutperturb}{\emph{User-Local-Out-Perturb}}
\newcommand{\fedsgd}{\emph{FedSGD}}
\newcommand{\fedavg}{\emph{FedAvg}}
\newcommand{\individual}{\emph{Individual}}
\newcommand{\localgroupdp}{\emph{LocalGDP}}
\newcommand{\higradavg}{\emph{HiGradAvgDP}}
\newcommand{\centralsubdp}{\emph{CentralSubDP}}
\newcommand{\localsubdp}{\emph{LocalSubDP}}

\newcommand{\ouralgo}{\emph{DecGDP}}
\newcommand{\ouralgobase}{\emph{DecGDP-Base}}
\newcommand{\ouralgofull}{\emph{DecGDP-Full}}

\newcommand{\femnistlarge}{\emph{FEMNIST-Large}}
\newcommand{\femnistsmall}{\emph{FEMNIST-Small}}

\newcommand{\noiseagg}{$\mathscr{G}$}
\newcommand{\noiseshuffler}{$\mathscr{S}$}
\newcommand{\cD}{\mathcal{D}}
\newcommand{\cM}{\mathcal{M}}
\newcommand{\cA}{\mathcal{A}}
\newcommand{\cS}{\mathcal{S}}
\newcommand{\bS}{\mathbb{S}}
\newcommand{\cC}{\mathcal{C}}
\newcommand{\cF}{\mathcal{F}}
\newcommand{\bE}{\mathbb{E}}

\newcommand{\pllm}{\textsf{PermLLM}}
\newcommand{\prag}{\textsf{PermRAG}}
\newcommand{\relv}{$relv$}
\newcommand{\multilora}{\emph{Activate}}
\newcommand{\multiloramerge}{\emph{Merge}}
\newcommand{\multiloraunion}{\emph{Union}}

\SetKwInput{KwInput}{Parameters}
\SetKwInput{KwOutput}{Output}

%% file: abstract.tex
\begin{abstract}

  In enterprise settings, organizational data is segregated, siloed
  and carefully protected by elaborate access control frameworks.
  These access control structures can completely break down if an LLM
  fine-tuned on the siloed data serves requests, for downstream tasks,
  from individuals with disparate access privileges.  We propose
  \emph{Permissioned LLMs (\pllm)}, a new class of LLMs that
  superimpose the organizational data access control structures on
  query responses they generate.  We formalize abstractions
  underpinning the means to determine whether access control
  enforcement happens correctly over LLM query responses.  Our
  formalism introduces the notion of a \emph{relevant response} that
  can be used to prove whether a \pllm\ mechanism has been implemented
  correctly.  We also introduce a novel metric, called \emph{access
  advantage}, to empirically evaluate the efficacy of a
  \pllm\ mechanism.  We introduce three novel \pllm\ mechanisms that
  build on Parameter Efficient Fine-Tuning to achieve the desired
  access control.  We furthermore present two instantiations of access
  advantage--(i) \emph{Domain Distinguishability Index (DDI)} based on
  Membership Inference Attacks, and (ii) \emph{Utility Gap Index
  (UGI)} based on LLM utility evaluation.  We demonstrate the efficacy
  of our \pllm\ mechanisms through extensive experiments on five
  public datasets (GPQA, RCV1, SimpleQA, WMDP, and PubMedQA), in addition to
  evaluating the validity of DDI and UGI metrics themselves for
  quantifying access control in LLMs.
  
\end{abstract}

%% file: intro.tex
\section{Introduction}
\label{sec:intro}

Large Language Models (LLMs), due to their unprecedented natural
language processing capabilities, are being adopted in a vast range of
applications across the entire computing
industry~\citep{kaddour23,zhao25}.
%% Soon enough LLMs will be routinely
%% used by individuals, both personally and as a part of an organization,
%% for everyday tasks that include information retrieval, content
%% creation, comprehension, summarization, planning, code generation and
%% debugging, education, classification and recommendation, to name a
%% few. In fact,
The day may not be too far off when LLMs become the primary interface
to a large swath of computing and information extraction tasks.  In
this paper, we focus on enterprise settings where LLMs are used to
perform a wide variety of computing tasks using organization-wide
data.  Using LLMs that have a wide purview over organizational data
brings massive troves of information and utility, including the
ability to combine learnings from disparate information silos of the
organization, to the finger tips of individuals in the organization.
%This is certainly an exciting prospect.  
However,
making all the learnings from organizational data available to any
individual who can query the LLM becomes a critical security
challenge: Organizations have access control structures and
hierarchies that tightly control information flow to and from
individuals within them. Information access via LLMs, if not carefully
controlled, risks undermining the existing access control structures
%is likely to break down those access control structures
and hierarchies.

As an example, consider government agencies using LLMs for a multitude
of tasks.  The data in government agencies is
typically segregated in multiple ``clearance levels'' and users can
access just the data they have access privileges for~\citep{nistsp20}.  Any other
agency data is inaccessible to the users.  An LLM trained on this
agency-wide data can leak privileged information to unauthorized
users, thus breaking the agency's access control framework that works
on the raw data.  Another example is that of role-based access
control~\citep{ferraiolo99,ferraiolo07}: Consider a health clinic
setting, where individuals performing different ``roles'' (doctors,
nurses, technicians, administrative staff, patients, etc.)  interact
with an LLM to perform many tasks.  The roles of the users determine
what part of the clinic-wide data they should have access to.  An LLM
trained on the clinic-wide data can be easily tricked into leaking
information to unauthorized users.

% The access control problem with LLMs has started to show up in the LLM
% research literature.
Research proposals to build system prompts that instruct an LLM to
control what information is generated in the output can help curb some
leakage of sensitive information to unauthorized
users~\citep{chen23a,liu25}.  However, they do not achieve absolute
security, and clever jailbreaking prompts~\citep{NEURIPS2023_fd661313,
  ShangYYSFZJ24, liu2024jailbreaking, NEURIPS2024_70702e8c} can be
used to overrule these system prompts.  A recent work proposes tagging
LLM queries with encrypted access credentials to authenticate users
and block unauthorized queries~\citep{chan25}.  This is a good start,
but it lacks the flexibility needed to enable access to disparate
learnings from the LLM for different users based on their access
credentials.  We discuss access control problems and solutions for
agentic systems and Retrieval Augmented Generation (RAG)
systems~\citep{lewis20} in~\autoref{sec:discussion}.

\begin{figure}[tb]
    \centering
    \includegraphics[width=0.9\linewidth]{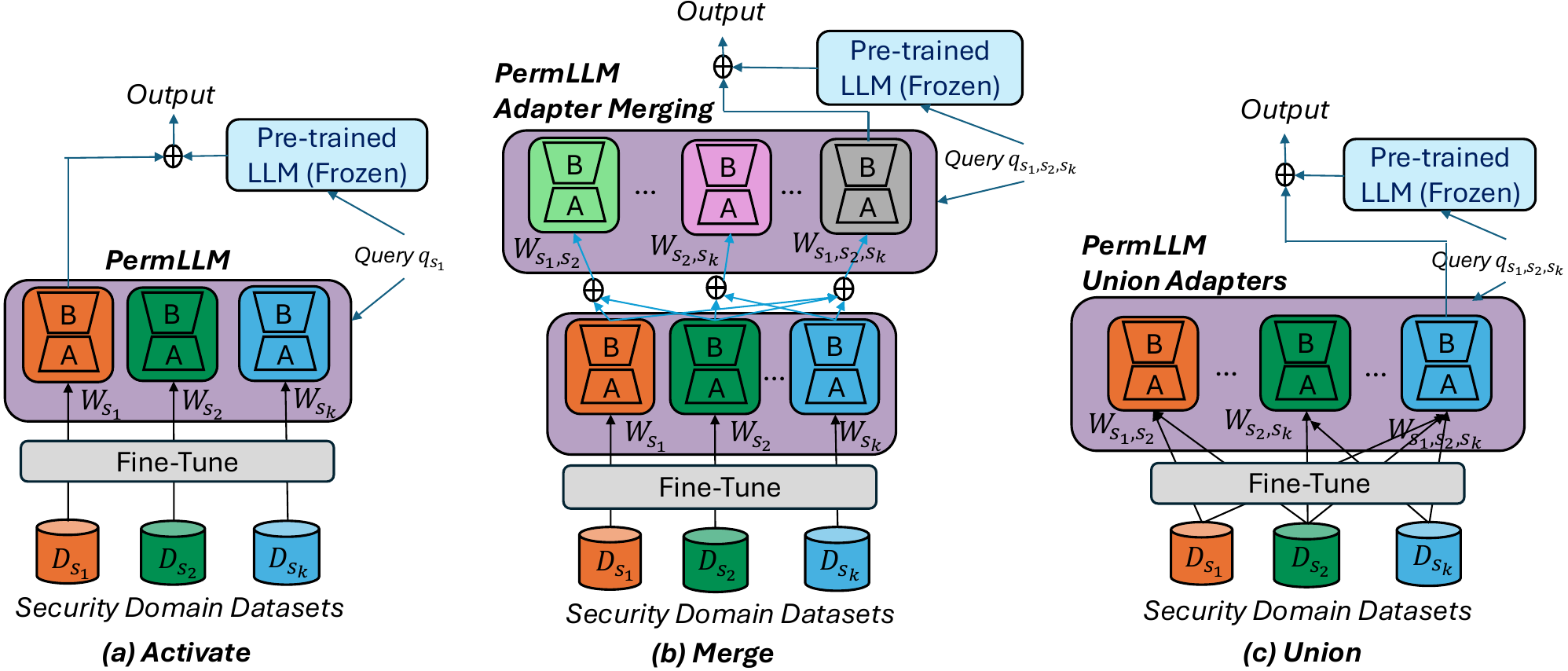}
    \caption{We propose three types of Permissioned LLM (\pllm)
      mechanisms. (a) \multilora: that has one-to-one mapping between
      the security domains and PEFT adapters. When a user queries the
      model, the mechanism activates the relevant adapter(s). (b)
      \multiloramerge: merges subsets of relevant PEFT adapters to serve the
      users that have access to multiple security domains. (c)
      \multiloraunion: trains adapters on the unions of various
      security domains, and at the inference phase the relevant PEFT
      adapter is activated to serve a user query that requires access
      to multiple security domains.}
    \label{fig:permllm}
  \end{figure}

This paper focuses on the access control problem for LLMs when they
are tuned on data coming from a multitude of data silos.  The
challenge here is to \emph{guarantee} that users who do not have
access to specific data silos cannot retrieve information from those
silos by sending carefully crafted queries to the LLMs tuned on data
from those silos. A recent work \citep{fleshman2025adapterswap} took
an initial step in this direction, but lacks the formal framework to
evaluate the access control. Moreover they only explore one type of
mechanism. As a security problem, access control is \emph{absolute}-- 
you either achieve access control or not, hence probabilistic solutions (e.g. Differential
Privacy~\citep{dwork06}) are not satisfactory.

\paragraph{Contributions.} In this paper, we comprehensively study the problem of access control in LLM fine-tuning. More specifically: (i) We formalize the notion of
access control mechanism in LLMs in terms of the \emph{relevance} of
responses generated by an LLM to the raw data the users have access
to. We use the notion of \emph{security domains} in our formalism. Our
formalism of response relevance can be used to prove correctness of
access control mechanisms. We also propose a novel metric called
\emph{access advantage} that helps us empirically quantify the
effectiveness of an access control mechanism on LLMs (\autoref{sec:problem_setup}).  
%Our additional contributions include: 
(ii) We present three new \pllm\ fine-tuning mechanisms (see
\autoref{fig:permllm}), based on Parameter Efficient Fine-Tuning
(PEFT)~\citep{hu21,xu23} (\autoref{sec:algos}).  (iii) We introduce
two novel instances of our access advantage metric, \emph{Domain
Distinguishability Index (DDI)} and \emph{Utility Gap Index (UGI)}, as
tools to audit access control enforcement via an adversarial gaming
setting (\autoref{sec:auditing}). (iv) We empirically evaluate our
access control mechanisms on two LLMs (Mistral-0.1-7B and
Llama-3.1-8B) using five different data sets: GPQA~\citep{gpqa},
RCV1~\citep{rcv1}, SimpleQA~\citep{simpleqa}, WMDP~\citep{wmdp}, and PubMedQA~\citep{jin-etal-2019-pubmedqa}
(\autoref{sec:exp}).  Our evaluation shows the effectiveness of our
access advantage metrics in assessing whether a proposed access
control mechanism for LLMs is enforcing data protection correctly.

\begin{comment}
Our key
contributions are as follows:
\begin{itemize}
    \item We propose \emph{Permissioned LLMs (\pllm{}s)} -- LLMs that
      provide access control security guarantees for the data they
      have ingested through tuning, and even through
      contexts/prompts. We theoretically define the notions of access
      control and access advantage, and show that our
      \pllm\ mechanisms provide access control
      (\autoref{sec:problem_setup}).
    \item We present three new \pllm\ fine-tuning mechanisms (see
      \autoref{fig:permllm}), based on Parameter Efficient Fine-Tuning
      (PEFT)~\citep{hu21,xu23} (\autoref{sec:algos}).
    \item We introduce two novel instances of our access advantage
      metric, \emph{Domain Distinguishability Index (DDI)} and
      \emph{Utility Gap Index (UGI)}, as tools to audit access control
      enforcement via an adversarial gaming setting
      (\autoref{sec:auditing}).
    \item We empirically evaluate our access control mechanisms on two
      LLMs (Mistral-0.1-7B and Llama-3.1-8B) using four different data
      sets: GPQA~\citep{gpqa}, RCV1~\citep{rcv1},
      SimpleQA~\citep{simpleqa}, and WMDP~\citep{wmdp}
      (\autoref{sec:exp}).
\end{itemize}
\end{comment}

%% file: formalism.tex
\section{Formalizing Access Control in LLMs}
\label{sec:problem_setup}

\subsection{Basic Setup and Notation}

We define a \emph{security domain} (henceforth called ``domain'' for
brevity) as a collection of data records that require identical
credentials for access (e.g. files with the same group in their access
control lists).  We consider settings where pretrained LLMs (such as
Llama and Mistral models) are fine-tuned over data from different
domains with an added constraint -- responses to inference time
queries will be generated from learnings on data coming from just the
domains the user has access to.  This added constraint is
enforced via access control mechanisms that govern how the LLM uses
data from different domains.

Consider a universe of $n$ different domains $\bS = \bigcup_{i=1}^n
\{s_i\}$, and a training data set consisting of data from these
domains $D = \bigcup_{i = 1}^n D_{s_i} \sim \cD_{s_i}$ (here $D_{s_i}$
is a data set sampled from data distribution $\cD_{s_i}$ of domain
$s_i$).  Let $f_{D}$ be the LLM tuned using data set $D$.  Let $W$ be
the set of $f_{D}$'s parameters.  Model fine-tuning \emph{changes}
values of a subset of $W$.  We say that a domain $s_i$ \emph{affects}
a subset of parameters $W_{s_i} \subseteq W$ if data from $D_{s_i}$ is
used to change parameters $W_{s_i}$ during model fine-tuning (unless
stated otherwise, the terms ``affect'' and ``affected'' mean this
relation between $s_i$ and $W_{s_i}$ in the rest of the paper).  We
define $\cM$ as an access control mechanism that dictates the mapping
of domain $s_i$ to parameters $W_{s_i}$ via the affects relation.  We
say that a LLM fine-tuned using data set $D$ is \emph{permissioned}
(\pllm), denoted as $f_{D}^{\cM}$, if it uses the access control
mechanism $\cM$ to map its parameters $W$ to a multitude of domains
from $\bS$, where each domain $s_i$ affects parameters $W_{s_i}
\subseteq W$.  Operationally, during fine-tuning, $\cM$ specifies
which set of model parameters $W_{s_i}$ are tuned for a given domain
$s_i$ (see \autoref{sec:algos} for more details). By the same token,
during inference, $\cM$ can specify which set of model parameters
should be used to answer a query based on the user's access
credentials.

We assume a setting where the \pllm\ $f_{D}^{\cM}$ resides in an
enclosing system $\cS$ that authenticates users who send queries
to $f_{D}^{\cM}$.  $\cS$ determines the user $u$'s access
credentials $cred_u$ and calls \texttt{authenticate($cred_u$)} that
takes user credentials $cred_u$ and maps them to a subset of 
domains $S_u$ that $u$ can access.  $S_u$ is maintained by $\cS$
and is never exposed to user $u$.  This process ensures $u$ cannot
arbitrarily change $S_u$.  Each of user $u$'s subsequent query $q$
to $f_{D}^{\cM}$ is annotated with $S_u$ by $\cS$.  $\cM$
determines the model parameters $W_{S_u}$ used to generate a response
$r_{S_u}$ to $q$, where $W_{S_u} = \bigcup_{s \in {S_u}}W_{s}$.

\subsection{Definitions}

\begin{definition}[Relevant Response]
  Given a \pllm\ $f_{D}^{\cM}$, a query $q$ from user $u$, and the set $S_u$
  of domains $u$ has access to, let $r = f_{D}^{\cM}(q)$
  be the response of $f_{D}^{\cM}$ to query $q$.  Response $r$ is
  said to be \emph{relevant} to $S_u$ (i.e., $r = r_{S_u}$) if
  $f_{D}^{\cM}$ uses parameters $W_{S_u}$ (in addition to any domain-agnostic model parameters) to generate $r$.
\label{def:relevant-response}
\end{definition}

We say that an access control mechanism $\cM$ is correctly
enforced on \pllm\ $f_{D}^{\cM}$ \emph{iff} every response $r$
generated for every user $u$'s query $q$ is relevant to $S_u$.

The above definition of relevant response helps us formally determine
if a proposed access control mechanism $\cM$ is algorithmically
correct. We however require an empirically quantifiable metric to
determine if the implementation (and the algorithm by extension) of
$\cM$ is correct. This is particularly important for auditing. To that
end, we propose a new metric called \emph{response relevance score},
$relv(f_D^\cM(q),S_u)$, which quantifies the information gained on
data in the domain set $S_u$ by
observing responses to queries generated using model parameters
$W_{S_u}$ affected by domains of $S_u$.  \relv\ is expected
to be higher when $q \sim \cD_{S_u}$ (i.e., $q$ is related to domain set $S_u$), compared to when $q \not\sim
\cD_{S_u}$.

We restrict the domain of \relv\ to the real number interval $[0,1]$,
where $1$ is the best expected score for relevance. \relv\ itself can
be represented by another empirical metric such as prediction
accuracy, or logits for the expected response. However, given that
LLMs (and ML models in general) are generalization engines, in
practice we expect \relv\ to be less than $1$.  This restriction can
be effectively addressed by measuring \relv\ for domains that
the user has access to and comparing it to \relv\ for domains
that the user does not have access to.  We call this the \emph{access
advantage}.

\begin{definition}[Access Advantage]
  Given \pllm\ $f_D^\cM$ trained over data set $D$ consisting of data
  from domains $\bS = \bigcup_{i=1}^n \{s_i\}$, with access control
  mechanism $\cM$, a subset of domains $S_u \subseteq
  \bS$, $f_D^\cM$ achieves \emph{$\alpha$-access advantage}
  w.r.t. $S_u$ if:
  \begin{equation*}
    \bE_{q \sim \cD_{S_u}, S_v \subseteq
    \bS; S_u \cap S_v = \phi}
    \big[ relv(f_D^\cM(q), S_u) \circleddash
      relv(f_D^\cM(q), S_v) \big] \ge \alpha
  \end{equation*}
  where $relv()$ is the response relevance score on the corresponding
  domain subset ($S_u$ or $S_v$), $\circleddash$ is a
  ``difference'' operator specific to the access control assessment
  method (e.g., subtraction), and $\alpha$ is an advantage threshold
  that lies in the range [0,1].
  \label{def:access-adv}
\end{definition}

The access advantage metric relies on the assumption that $f_D^\cM$
performs significantly better on domains user $u$ has access
to compared to domains $u$ does not have access to.  In other
words, $f_D^\cM$ should have explicit segregation between the
different domains, as dictated by $\cM$. The existing
approaches to model fine-tuning fail to achieve this goal as the model
is traditionally trained on all the domains without any
built-in domain segregation mechanism.
To the best of our knowledge, no prior work on LLM and
privacy formally tackles this problem of access control through explicit domain
segregation.  We next propose novel mechanisms to achieve domain
segregation in \autoref{sec:algos} and propose empirical metrics to
evaluate the access control protocols in \autoref{sec:auditing}.

We believe access advantage is a critical metric for auditors to
determine if an access control mechanism is truly achieving the
segregation of domains as intended.  Hence it is in the
auditor's best interest to ensure that $S_u \cap S_v = \phi$.
Access advantage can diminish significantly when $S_u \cap S_v \neq
\phi$, leading to incorrect conclusions about the efficacy of the
access control mechanism.

To the best of our knowledge, prior works on retrieval
augmented generation (RAG) based LLM deployments do not explicitly tackle the
problem of measuring effectiveness of access control mechanisms
formally or empirically.  Our formalism of relevant response and
access advantage extends to RAG systems as well, closing that gap in
formalism and empirical evaluation of access control protocols.  While
a thorough evaluation of access control for RAG based systems is
outside the scope of this paper, a more detailed analysis of conditions
for formal correctness of access control in RAG systems appears
in \autoref{sec:rag-formalism}.

\subsection{Auditing Access Control}\label{sec:threat_model}

To evaluate the access control mechanisms, we consider a classic
adversarial game between the system $\cS$\ enclosing the model
$f_D^\cM$ and the auditor $\cA$. We give $\cA$\ the ability to choose
domain access by emulating an end user, send arbitrary queries to the
model via $\cS$\ and observe the responses. $\cA$\ can replay the game
multiple times as different users to conclude if the access control is
correctly implemented.

\paragraph{Audit Game.} The formal game between auditor $\cA$ and system $\cS$\ is as follows:
\begin{enumerate}[itemsep=2pt,   % vertical space between items
	topsep=2pt,    % space before/after the list
	parsep=0pt,    % space between paragraphs in an item
	partopsep=0pt] % extra space when list starts a new part
  \item Auditor $\cA$ chooses domain set $S_u$ and emulates user $u$. $\cA$ sends user credentials $cred_u$ and query $q \sim \cD_{S_u}$ to system $\cS$.
  \item $\cS$ verifies the user credential $cred_u$ and sends back the model response $f_D^\cM(q)$ to $\cA$.
  \item $\cA$ chooses domain set $S_v$ such that $S_v \cap S_u = \phi$ and emulates user $v$. $\cA$ sends user credentials $cred_v$ and the same query $q \sim \cD_{S_u}$ to $\cS$.
  \item $\cS$ verifies the user credential $cred_v$ and sends back the model response $f_D^\cM(q)$ to $\cA$.
  \item $\cA$ concludes the access control mechanism is correctly implemented if the access advantage $|relv(f_D^\cM(q), S_u) \circleddash
      relv(f_D^\cM(q), S_v)| \ge \alpha$.
\end{enumerate}

Note that the auditor $\cA$ has superuser privileges to choose arbitrary domain access unlike an ordinary user. This is by design to allow the auditor to evaluate the correctness of the claimed access control while still following the protocol of querying the model as a benign user. Detailed instantiations of this adversarial game for different suites of access advantage metrics are discussed in \autoref{sec:audit_games}.

%% file: algos.tex
\section{Permissioned LLM Mechanisms}\label{sec:algos}

We rely on Parameter Efficient Fine-Tuning (PEFT)~\citep{hu21,xu23} to
obtain model parameter segregation for domains.  We focus on the LoRA
PEFT adapter~\citep{hu21}, however our mechanisms seamlessly apply to
other types of adapters~\citep{houlsby19,xu23}.  The three mechanisms
we describe ensure that domain data is steered to train select LoRA
adapters.  Each domain has a unique identifier (domain Id).  Our
access control mechanism builds a map between domains and LoRA
adapters within the \pllm{}'s metadata.  The map is used to steer all
examples from a domain to the corresponding adapter/s for training.
This map is also used to steer queries to the correct LoRA adapters at
inference time.

\paragraph{One LoRA per Security Domain}
For our base mechanism called \multilora, we assume that users have
access to at most one domain.  \autoref{fig:permllm}(a) depicts our
base mechanism that performs a simple 1-1 mapping between domains and
LoRA adapters.  We assume that the number of domains is known
beforehand, and use that knowledge to instantiate corresponding number
of LoRA adapters.  During training, each minibatch is sampled from one
domain, and the domain's Id is used to select the LoRA adapter to
train.  At inference time, a user's query is annotated with the domain Id
the user has access to.  This domain Id is used to \emph{activate} the
LoRA adapter for the corresponding domain.

\paragraph{Merging LoRA Adapters for Security Domain Groups}
In many application settings, users have access to data from multiple
domains.  For queries coming from such users, our \multilora\ enables
all corresponding LoRA adapters, whose activations are averaged at
inference time. We however found that activations from different LoRA
adapters tend to disruptively interfere with each other resulting in
catastrophic performance degradation beyond two domains. We leave
further refinement of activation space steering~\citep{shen25,yin24}
to future work.  In our second mechanism,
\multiloramerge\ (\autoref{fig:permllm}(b)), we adopt the LoRA adapter
merging strategy for users with access to multiple
domains~\citep{stoica24,yadav23,yu24,zhao24}.  We experimented with
several LoRA merging algorithms including TIES~\citep{yadav23} and
DARE~\citep{yu24}, but eventually favored the SVD
approach~\citep{stoica24} because of its better performance and
stability in the context of LoRA merging. We assume that the combination of domains that users may have access to are known beforehand.  Thus, after training LoRA
adapters for individual domains, we can merge them for those exact
domain combinations.  Correspondingly, our domain-LoRA adapter map is
updated with the domain IDs and the merged LoRA adapters.  These new
mappings are used at inference time to activate the correct merged LoRA
adapters.
We found that adapter merging is more robust to cross-adapter
interference than activation merging.

\paragraph{Training a LoRA per Combination of Security Domains}
Although \multiloramerge\ is better than activation space merging of
multiple LoRA adapters, we observed that it also leads to model performance
degradation with increasing number of merged adapters.  As a result,
we explored another simple alternative, \multiloraunion, which
\emph{trains} a LoRA adapter on data from each unique combination of
domains users have access to.
\multiloraunion\ indeed delivers the best performance in all our
mechanisms.  However, it comes at the cost of significantly greater
tuning time compute -- a domain can occur in numerous combinations of
domains (e.g. in~\autoref{fig:permllm}(c), data $D_{s_2}$ gets used in
the training set of all three LoRA adapters).  Furthermore, data sets
containing large number of domains presents the added challenge of an
exponential blow up in domain combinations (at most $2^n$).  However,
we believe the number of combinations present in a real-world setting
will be much smaller than that upper bound.

The careful mapping of domains (or groups of domains) to the correct
LoRA adapters, and steering of training examples from domains to the
corresponding LoRA adapters ensures precise parameter segregation for
domains.  Our assumption that users cannot tamper with their access
credentials at inference time further aids the \pllm{}'s enclosing system
to determine the correct set of domains corresponding to a query.  The
query steering that happens through the \pllm\ using domain IDs
\emph{guarantees} that all responses generated by the \pllm\ are
\emph{relevant} to the user's domains.  Furthermore, the responses are
not generated using LoRA adapters that were trained using data from
domains that the user does \emph{not} have access to. 
Response relevance for all responses implies correctness of our
\pllm\ access control mechanisms.  Our proof appears
in~\autoref{sec:algs-formalism}.

%% file: audit.tex
\section{Auditing Access Control in Permissioned LLM Mechanisms}
\label{sec:auditing}

We now introduce two novel instantiations of our \emph{access
advantage} metric (\sref{Definition}{def:access-adv})—Domain
Distinguishability Index (DDI) and Utility Gap Index (UGI)—that
quantify access control efficacy independently of any particular
system design. We show how these metrics fit into the framework for
empirically assessing access control mechanisms in \pllm{}s through
adversarial audit games in \autoref{sec:audit_games}. These metrics
are in [0,1] range with higher values denoting better access control.

\subsection{Metric 1: Domain Distinguishability Index (DDI)}

In traditional privacy evaluations, membership inference attacks (MIAs) leverage a sampled member data set (from the target model’s training set) and a sampled non-member data set to assess privacy leakage~\citep{jayaraman2021revisiting,shokri2017membership}: the more accurately an adversary separates and classifies samples as members or non-members, the higher the privacy risk. Analogously, we adopt this MIA framework for access control assessment to distinguish security domains. Specifically, for any security domain set $S_i$, the auditor holds samples from $S_i$'s training data (member set) and samples from $S_j$'s training data (non-member set), where $S_j \cap S_i = \phi$.  The auditor then evaluates how successfully it can distinguish the member set from the non-member set when the \pllm\ is activated for $S_i$. This evaluation occurs for all security domains, giving us an aggregate access advantage, which we call Domain Distinguishability Index (DDI).

\begin{definition}[Domain Distinguishability Index (DDI)]\label{def:ddi}
	For a domain universe $\bS$ consisting of $n$ security domains, let $f_D^\cM$ denote the \pllm\ trained on data $D$ from all security domains with access control mechanism $\cM$. For each ordered pair of domain sets $(S_i \subseteq \bS, S_j \subseteq \bS)$ with no overlap (i.e.,$S_i \cap S_j = \phi$), let
	$O^{(S_i,S_j)} = O(f_D^\cM(q)|S_i, f_D^\cM(q)|S_j); \forall q \sim \cD_{S_i}$ be the output of a membership inference
	oracle $O$. For a given membership inference metric $m(\cdot)$, the DDI is defined as:
	$\mathrm{DDI}(m) = 
		\bE_{S_i \subseteq \bS, S_j \subseteq \bS} \big[m\bigl(O^{(S_i,S_j)}\bigr)\big]$, where $\bE$ is the expectation over all domain sets.
\end{definition}

We also report the standard deviation of
$m\bigl(O^{(S_i,S_j)}\bigr)$ across all domain set pairs to capture
variability.  By~\ref{def:access-adv}, DDI can be viewed as an
access advantage metric, where the response relevance score $relv$ for $S_i$ on query $q$, $relv(f_D^\cM(q),S_i)$, is a binary value on whether the membership inference oracle $O$'s output is above a membership threshold.  The difference operator $\circleddash$ is the MIA method specific composition of response relevance for all the samples in the member and non-member sets.

We use AUC-ROC and TPR@$(low)$FPR, as instantiations of DDI, where higher scores indicate stronger enforcement, as $S_i$-specific responses become more distinguishable. See Appendices~\ref{app:MIA_metrics} and~\ref{app:existing_MIAs} for details on MIA evaluation metrics and an overview of existing MIAs against LLMs.

A higher DDI indicates more robust separation between security
domains. In our evaluations, we employ state-of-the-art MIAs
for LLMs, including Loss~\citep{yeom2018privacy},
Zlib~\citep{carlini2021extracting}, Mink~\citep{shi2023detecting},
Mink++~\citep{zhang2024min}, Reference~\citep{carlini2021extracting}
attacks.

\subsection{Metric 2: Utility Gap Index (UGI)}

The UGI metric measures the drop in model utility on the target domain's data when a different domain's adapter is activated in \pllm\ instead of the target domain.

\begin{definition}[Utility Gap Index (UGI)]\label{def:ugi}
	Let $U(\cdot)$ be a chosen utility metric and for a domain set pair $(S_i \subseteq \bS, S_j \subseteq \bS)$ with no overlap (i.e.,$S_i \cap S_j = \phi$), $\mathrm{UtilityGap}^{(S_i,S_j)}(U) = |U(f_D^\cM(q)|S_i) - U(f_D^\cM(q)|S_j)|; \forall q \sim \cD_{S_i}$.  The UGI aggregates utility gaps across all ordered domain set pairs:
		$\Delta_{U}
		\;=\;
		\bE_{S_i \subseteq \bS, S_j \subseteq \bS}
		\big[\mathrm{UtilityGap}^{(S_i,S_j)}(U)\big]$, where $\bE$ is the expectation over all domain sets.
\end{definition}

By~\ref{def:access-adv}, UGI is also an instantiation of the access advantage
metric in which the relevance score for security domain set $S_i$ on query $q$ is the
utility value itself, $\mathit{relv}\!\bigl(f_{D}^{\cM}(q),S_i\bigr)=U\!\bigl(f_{D}^{\cM}(q)|S_i\bigr)$,
and the operator $\circleddash$ computes the absolute
difference of those relevance scores across the sampled queries.

A larger UGI indicates that enforced access controls yield more
pronounced—and thus more easily perceivable—differences in response
quality between security domains.  As with DDI, we also report the
standard deviation across pairs to characterize variability. We evaluate the utility gaps w.r.t. Bleurt Score ($\Delta_{bluert}$), Bert F1-Score ($\Delta_{bert}$), Sacrebleu Score ($\Delta_{bleu}$) and Verbatim Accuracy ($\Delta_{acc}$) for our UGI metrics in \autoref{sec:exp}. More details on these metrics can be found in Appendix \autoref{sec:utility_eval}.

%% file: eval.tex
\section{Experimental Evaluation}\label{sec:exp}

For our experiments, we fine-tune
Llama-3.1-8B and Mistral-0.1-7B pretrained models on five datasets
covering multiple distinct security domains (henceforth called
\emph{domains}), where we fine-tune a separate LoRA adapter for each
domain. Details about the model hyperparameters can be found in Appendix \autoref{sec:models}. The data sets we use in our experiments are WMDP~\citep{wmdp}, GPQA~\citep{gpqa}, SimpleQA~\citep{simpleqa},
RCV1~\citep{rcv1}, and PubMedQA~\citep{jin-etal-2019-pubmedqa}. \autoref{tab:datasets_abridged} shows the brief
data set details. More details on the data sets and generalization gaps can be found in Appendix \autoref{sec:data_sets}. Appendix \autoref{sec:utility_eval} discusses the utility evaluation of all our models.

\begin{table}[tb]
    \centering \small
    \caption{Data Set Details.}
    \begin{tabular}{lrrrrr}
        \toprule
         & WMDP & GPQA & SimpleQA & RCV1 & PubMedQA \\
        \midrule
        Data Set Size (Train / Test)  & 2936 / 732 & 360 / 88 & 4089 / 1018 & 45622 / 22811 & 200000 / 11269 \\
        Number of Security Domains  & 3    & 3   & 10   & 4   & 10 \\
        \bottomrule
    \end{tabular}
    \label{tab:datasets_abridged}
\end{table}

\subsection{Evaluating Access Control}
\label{sec:access_control}

Our approach achieves comparable model utility to existing approaches
of fine-tuning (see discussion in \autoref{sec:utility_eval}), in
addition to providing access control. Here we will empirically
evaluate the effectiveness of our access control using a suite of
metrics. We will first consider the case where the user has access to
only one domain and report the access control results in
\autoref{sec:single_active_domain}. Next in
\autoref{sec:multi_active_domain}, we will consider the case where the
user has access to multiple domains. For comparison, we also include an evaluation of a prompt-based access control baseline in \autoref{sec:prompt_based} but find it to be ineffective.

\subsubsection{Single Active Domain}
\label{sec:single_active_domain}

In Section~\ref{sec:auditing}, we proposed an \emph{adversarial audit
framework} for empirically assessing access control in \pllm{}s.  We
introduced two concrete instantiations of the general \emph{access
advantage} metric: the Domain Distinguishability Index (DDI) and the
Utility Gap Index (UGI)~$\Delta_{U}$.  Although
\autoref{sec:algos} gives formal guarantees—each response is computed
solely from domains the user is authorized to access—we \emph{measure}
access control enforcement strength with DDI and UGI
($\Delta_{U}$) to confirm that the guarantees hold in practice,
which is necessary to verify correctness of \emph{implementations}.

\begin{figure}[tb]
    \includegraphics[width=\textwidth]{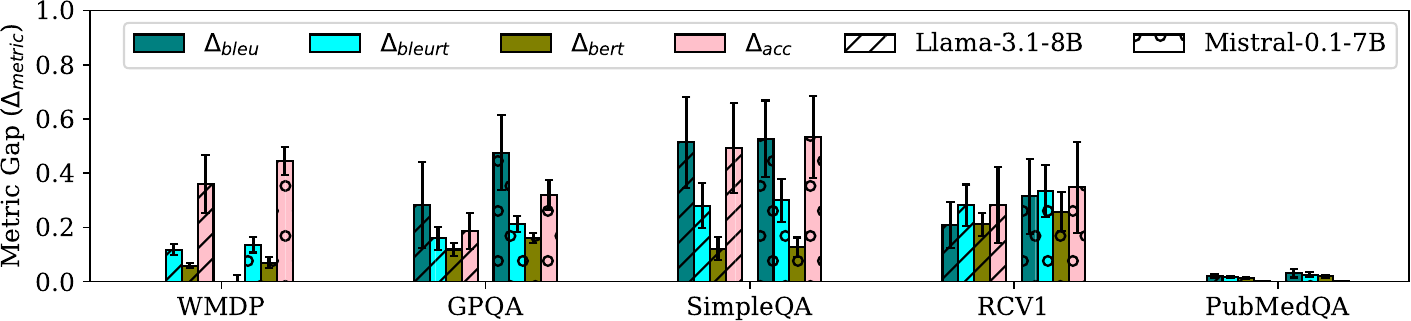}
    \caption{Utility Gap Index, $\Delta_U$ ($mean \pm std$) when user
      has access to one security domain.}
    \label{fig:utility_gap_act_1}
\end{figure}

Theoretically, $\Delta_{U}$ may reach~1.0, but empirically we
observe much smaller—yet substantial—access advantage gaps for four of the data sets
(\autoref{fig:utility_gap_act_1}).  These gaps are significantly
impacted by domain distributions and the strictness of the scoring
metric.  For example, SimpleQA exhibits the largest UGIs (up to
$\Delta_{blue}\!\approx\!0.50$ and $\Delta_{acc}\!\approx\!0.50$)
because it has the highest number of distinct domains (10 in total).
Moreover, we observe that $\Delta_{bleu}$ and $\Delta_{acc}$ have the
largest values as these metrics look for verbatim pattern matches,
thus requiring the model to memorize the nuances in the target
domain. On the other hand, $\Delta_{bleurt}$ and $\Delta_{bert}$ look
for approximate similarities, and hence are impacted by the
similarities across the domains. This suggests that the verbatim
  matching metrics, $\Delta_{bleu}$ and $\Delta_{acc}$, are better
  model utility metrics for measuring access advantage compared to the
  similarity based metrics $\Delta_{bleurt}$ and $\Delta_{bert}$. For
large data sets like RCV1, all the metrics achieve similar values as
the model begins to generalize more. While these values are not close
to 1, they still provide credence to the fact that the domains are
different and our access control protocol works as expected due to the
utility gaps. The access advantage threshold $\alpha$ is dependent on the type of utility metric: verbatim
matching metrics $\Delta_{bleu}$ and $\Delta_{acc}$ have higher threshold than similarity based metrics $\Delta_{bleurt}$ and $\Delta_{bert}$. For $\Delta_{acc}$ metric, $\alpha > 0.2$ is sufficient to infer that access control is happening correctly. PubMedQA is an exception where $\Delta_{U}$ values are close to zero; this is because the security domains are artificially obtained via k-means and hence have the same underlying data distribution.

\autoref{tab:attack-results} shows DDI values obtained from a suite of
state-of-the-art MIAs.  Across domain pairs, the access advantage
(distinguishability) scores approach $\alpha ~= 1.0$, indicating that
an external auditor can almost perfectly identify the active domain (even when the domain distributions are similar as in the case of PubMedQA).
Hence, even when UGI values fall significantly below $1.0$ because of
model generalization, the high DDI values show that access control in
\multilora\ still functions as intended.  This clearly suggests that
DDI is the better method for \pllm\ access control efficacy
evaluation.

\input{MIAs_results}

\subsubsection{Multiple Active Domains}\label{sec:multi_active_domain}
\begin{figure}[tb]
    \includegraphics[width=\textwidth]{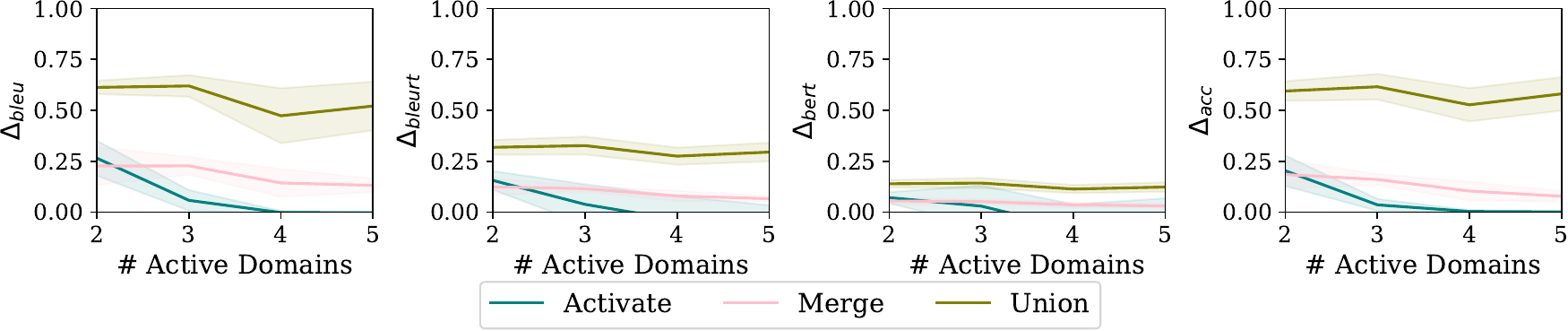}
    \caption{Utility Gap Index, $\Delta_U$ ($mean \pm std$) for
      Llama-3.1-8B models fine-tuned on SimpleQA when user has access
      to multiple security domains.}
    \label{fig:utility_gap_simpleqa_llama}
\end{figure}

As discussed earlier in \autoref{sec:algos}, we explore three methods
of combining knowledge from multiple domains the user has access to:
(a) activating all the domain-specific LoRA modules (\multilora), (b)
merging the LoRA modules (\multiloramerge), and (c) training separate
LoRA modules on the union of domains and using those for inference
(\multiloraunion). \autoref{tab:utility_gap_act_2} reports the UGI
($\Delta_{U}$) for these approaches when the user has access to two
domains for all the data sets. We note that WMDP and GPQA have only
three security domains, and hence activating any two domains always
lead to overlap when calculating $\Delta_{U}$ as per
\ref{def:ugi}. For these data sets, we calculate $\Delta_{U}$ on the
non-overlapping data. \multilora\ is computationally inexpensive but
suffers from considerable utility loss when compared to the previous
case of single domain. This is due to the high interference across the
multiple domains in the activation space, which is a known issue in
the multi-task learning literature~\citep{Zhao_2018_ECCV,
  vu2022spotbetterfrozenmodel,
  ostapenko2024modularllmsbuildingreusing}. The utility loss
suppresses the absolute $\Delta_{U}$ in our experiments.
As can be seen in \autoref{fig:utility_gap_simpleqa_llama},
\multiloramerge\ reduces the cross-domain interference, but still
suffers from utility loss.  Interestingly \multiloramerge\ achieves
even lower $\Delta_{U}$ than \multilora\ when combining two domains,
as shown in \autoref{tab:utility_gap_act_2}. Although it quickly
outperforms \multilora\ when the user has access to more than two
domains, the utility loss due to model merging
interference~\citep{stoica24,yadav23,yu24,zhao24} also results in
progressive degradation of $\Delta_{U}$ (see
\autoref{fig:utility_gap_simpleqa_llama}).
\multiloraunion\ retains $\Delta_{U}$ even beyond four domains, and
hence is the best choice when combining knowledge from several
domains. But this comes at the cost of more training-time computation
since new domain-specific modules have to be trained for the union of
domains, and there could be potential combinatorial blow-up of the
number of such combinations. As with the single active domain case, we observe close to zero utility gap on PubMedQA as the domains share the same data distribution. We observe similar results for
Mistral-0.1-7B model (see \autoref{fig:utility_gap_simpleqa_mistral}
in the appendix).

\begin{table}[tb]
    \centering
    \small
    \caption{Utility Gap Index ($\Delta_U$) for models with different
      approaches of combining domains when user has access to two
      domains. All reported values are $mean \pm std$ across domains.}
    \begin{tabular}{llcccccc}
        \toprule
        & Metric & \multicolumn{3}{c}{Llama-3.1-8B} & \multicolumn{3}{c}{Mistral-0.1-7B}\\
        & & \multilora & \multiloramerge & \multiloraunion & \multilora & \multiloramerge & \multiloraunion \\
        \midrule
        \multirow{3}{*}{\rotatebox[origin=c]{90}{WMDP}} 
        & $\Delta_{bleurt}$ & $0.09 \pm 0.01$ & $0.07 \pm 0.02$ & $0.11 \pm 0.02$ & $0.10 \pm 0.02$ & $0.08 \pm 0.03$ & $0.14 \pm 0.03$ \\
        & $\Delta_{bert}$   & $0.05 \pm 0.01$ & $0.03 \pm 0.01$ & $0.06 \pm 0.01$ & $0.05 \pm 0.01$ & $0.04 \pm 0.02$ & $0.07 \pm 0.02$ \\
        & $\Delta_{acc}$    & $0.27 \pm 0.07$ & $0.21 \pm 0.09$ & $0.34 \pm 0.11$ & $0.32 \pm 0.04$ & $0.25 \pm 0.07$ & $0.49 \pm 0.09$ \\
        \midrule
        \multirow{4}{*}{\rotatebox[origin=c]{90}{GPQA}} 
        & $\Delta_{bleu}$   & $0.15 \pm 0.06$ & $0.11 \pm 0.06$ & $0.51 \pm 0.07$ & $0.24 \pm 0.10$ & $0.17 \pm 0.10$ & $0.62 \pm 0.02$ \\
        & $\Delta_{bleurt}$ & $0.10 \pm 0.02$ & $0.06 \pm 0.02$ & $0.26 \pm 0.03$ & $0.14 \pm 0.06$ & $0.10 \pm 0.04$ & $0.32 \pm 0.02$ \\
        & $\Delta_{bert}$   & $0.07 \pm 0.02$ & $0.04 \pm 0.03$ & $0.18 \pm 0.02$ & $0.11 \pm 0.04$ & $0.08 \pm 0.03$ & $0.21 \pm 0.02$ \\
        & $\Delta_{acc}$    & $0.09 \pm 0.04$ & $0.05 \pm 0.02$ & $0.31 \pm 0.08$ & $0.16 \pm 0.07$ & $0.08 \pm 0.07$ & $0.51 \pm 0.04$ \\
        \midrule
        \multirow{4}{*}{\rotatebox[origin=c]{90}{SimpleQA}} 
        & $\Delta_{bleu}$   & $0.26 \pm 0.09$ & $0.23 \pm 0.09$ & $0.61 \pm 0.03$ & $0.30 \pm 0.13$ & $0.25 \pm 0.04$ & $0.61 \pm 0.08$ \\
        & $\Delta_{bleurt}$ & $0.16 \pm 0.05$ & $0.12 \pm 0.04$ & $0.32 \pm 0.04$ & $0.19 \pm 0.05$ & $0.14 \pm 0.02$ & $0.33 \pm 0.05$ \\
        & $\Delta_{bert}$   & $0.07 \pm 0.03$ & $0.05 \pm 0.02$ & $0.14 \pm 0.02$ & $0.08 \pm 0.03$ & $0.06 \pm 0.01$ & $0.14 \pm 0.03$ \\
        & $\Delta_{acc}$    & $0.20 \pm 0.07$ & $0.18 \pm 0.07$ & $0.59 \pm 0.05$ & $0.27 \pm 0.09$ & $0.21 \pm 0.03$ & $0.62 \pm 0.09$ \\
        \midrule
        \multirow{4}{*}{\rotatebox[origin=c]{90}{RCV1}} 
        & $\Delta_{bleu}$   & $0.05 \pm 0.03$ & $0.04 \pm 0.02$ & $0.16 \pm 0.09$ & $0.04 \pm 0.02$ & $0.01 \pm 0.03$ & $0.19 \pm 0.10$ \\
        & $\Delta_{bleurt}$ & $0.11 \pm 0.01$ & $0.07 \pm 0.03$ & $0.22 \pm 0.08$ & $0.08 \pm 0.01$ & $0.03 \pm 0.04$ & $0.22 \pm 0.08$ \\
        & $\Delta_{bert}$   & $0.08 \pm 0.01$ & $0.06 \pm 0.02$ & $0.16 \pm 0.04$ & $0.06 \pm 0.01$ & $0.03 \pm 0.05$ & $0.18 \pm 0.06$ \\
        & $\Delta_{acc}$    & $0.03 \pm 0.01$ & $0.04 \pm 0.04$ & $0.24 \pm 0.14$ & $0.02 \pm 0.02$ & $0.01 \pm 0.03$ & $0.26 \pm 0.15$ \\
        \midrule
        \multirow{4}{*}{\rotatebox[origin=c]{90}{\scriptsize PubMedQA}} 
        & $\Delta_{bleu}$   & $0.01 \pm 0.00$ & $0.00 \pm 0.00$ & $0.01 \pm 0.00$ & $0.01 \pm 0.00$ & $0.00 \pm 0.00$ & $0.01 \pm 0.01$ \\
        & $\Delta_{bleurt}$ & $0.01 \pm 0.00$ & $0.00 \pm 0.00$ & $0.01 \pm 0.00$ & $0.01 \pm 0.00$ & $0.00 \pm 0.00$ & $0.01 \pm 0.01$ \\
        & $\Delta_{bert}$   & $0.01 \pm 0.00$ & $0.00 \pm 0.00$ & $0.01 \pm 0.00$ & $0.01 \pm 0.00$ & $0.00 \pm 0.00$ & $0.01 \pm 0.00$ \\
        & $\Delta_{acc}$   & - & - & - & - & - & -  \\
        \bottomrule
    \end{tabular}
    \label{tab:utility_gap_act_2}
\end{table}

\input{DDI_Multi}

%% file: MIAs_results.tex
\begin{table}[ht]
	\centering
	\small
	\caption{DDI values with $m \in\{\text{AUC–ROC},\text{TPR@1\%FPR},\text{TPR@5\%FPR}\}$ for the different MIAs.  
		Mink and Mink++ are run with hyperparameter $k = 10\%$.  
		Entries are reported as $mean \pm std$ across security domains.}
	\begin{tabular}{llcccccc}
		\toprule
		 & MIA & \multicolumn{3}{c}{Llama-3.1-8B} & \multicolumn{3}{c}{Mistral-0.1-7B} \\
		 & & \textsf{auc-roc} & \textsf{tpr@1\%fpr} & \textsf{tpr@5\%fpr} & \textsf{auc-roc} & \textsf{tpr@1\%fpr} & \textsf{tpr@5\%fpr} \\
		\midrule
		\multirow{5}{*}{\rotatebox[origin=c]{90}{WMDP}} 
		& Loss                & $0.99 \pm 0.01$ & $0.93 \pm 0.10$ & $0.96 \pm 0.06$ & $1.00 \pm 0.00$ & $0.95 \pm 0.06$ & $0.99 \pm 0.01$ \\
		& ZLIB                & $0.98 \pm 0.03$ & $0.77 \pm 0.31$ & $0.85 \pm 0.21$ & $0.99 \pm 0.02$ & $0.85 \pm 0.25$ & $0.92 \pm 0.14$ \\
		& Mink 		          & $1.00 \pm 0.00$ & $0.95 \pm 0.07$ & $1.00 \pm 0.01$ & $1.00 \pm 0.00$ & $1.00 \pm 0.00$ & $1.00 \pm 0.00$ \\
		& Mink++ 	          & $1.00 \pm 0.00$ & $0.99 \pm 0.01$ & $1.00 \pm 0.00$ & $1.00 \pm 0.00$ & $1.00 \pm 0.00$ & $1.00 \pm 0.00$ \\
		& Ref                 & $0.99 \pm 0.01$ & $0.93 \pm 0.10$ & $0.96 \pm 0.06$ & $1.00 \pm 0.00$ & $0.95 \pm 0.08$ & $0.98 \pm 0.03$ \\
		\midrule
		\multirow{5}{*}{\rotatebox[origin=c]{90}{GPQA}} 
		& Loss                & $0.97 \pm 0.05$ & $0.81 \pm 0.26$ & $0.94 \pm 0.08$ & $0.98 \pm 0.03$ & $0.93 \pm 0.10$ & $0.95 \pm 0.07$ \\
		& ZLIB                & $0.95 \pm 0.04$ & $0.45 \pm 0.22$ & $0.77 \pm 0.15$ & $0.97 \pm 0.02$ & $0.57 \pm 0.24$ & $0.83 \pm 0.13$ \\
		& Mink	           	  & $0.99 \pm 0.01$ & $0.94 \pm 0.11$ & $0.98 \pm 0.03$ & $1.00 \pm 0.00$ & $0.98 \pm 0.02$ & $0.99 \pm 0.02$ \\
		& Mink++			  & $1.00 \pm 0.00$ & $1.00 \pm 0.01$ & $1.00 \pm 0.00$ & $1.00 \pm 0.00$ & $0.99 \pm 0.01$ & $1.00 \pm 0.00$ \\
		& Ref                 & $1.00 \pm 0.00$ & $0.97 \pm 0.04$ & $0.99 \pm 0.01$ & $1.00 \pm 0.00$ & $0.97 \pm 0.05$ & $0.99 \pm 0.02$ \\
		\midrule
		\multirow{5}{*}{\rotatebox[origin=c]{90}{SimpleQA}} 
		& Loss                & $0.98 \pm 0.03$ & $0.81 \pm 0.34$ & $0.90 \pm 0.25$ & $0.99 \pm 0.03$ & $0.81 \pm 0.32$ & $0.92 \pm 0.20$ \\
		& ZLIB                & $0.98 \pm 0.03$ & $0.80 \pm 0.33$ & $0.90 \pm 0.23$ & $0.99 \pm 0.03$ & $0.80 \pm 0.33$ & $0.91 \pm 0.20$ \\
		& Mink	              & $0.99 \pm 0.03$ & $0.80 \pm 0.32$ & $0.91 \pm 0.21$ & $0.99 \pm 0.03$ & $0.82 \pm 0.31$ & $0.92 \pm 0.20$ \\
		& Mink++			  & $0.98 \pm 0.03$ & $0.81 \pm 0.32$ & $0.91 \pm 0.21$ & $0.99 \pm 0.03$ & $0.82 \pm 0.31$ & $0.92 \pm 0.21$ \\
		& Ref                 & $0.98 \pm 0.04$ & $0.78 \pm 0.36$ & $0.90 \pm 0.25$ & $0.98 \pm 0.03$ & $0.79 \pm 0.36$ & $0.90 \pm 0.24$ \\
		\midrule
		\multirow{5}{*}{\rotatebox[origin=c]{90}{RCV1}} 
		& Loss                & $0.99 \pm 0.01$ & $0.86 \pm 0.21$ & $0.97 \pm 0.06$ & $0.99 \pm 0.02$ & $0.85 \pm 0.24$ & $0.96 \pm 0.09$ \\
		& ZLIB                & $0.93 \pm 0.07$ & $0.71 \pm 0.26$ & $0.81 \pm 0.18$ & $0.94 \pm 0.08$ & $0.73 \pm 0.28$ & $0.83 \pm 0.19$ \\
		& Mink	         	  & $1.00 \pm 0.01$ & $0.94 \pm 0.10$ & $0.98 \pm 0.03$ & $0.99 \pm 0.01$ & $0.88 \pm 0.18$ & $0.98 \pm 0.05$ \\
		& Mink++			  & $1.00 \pm 0.00$ & $0.97 \pm 0.05$ & $0.99 \pm 0.01$ & $1.00 \pm 0.01$ & $0.96 \pm 0.06$ & $0.99 \pm 0.02$ \\
		& Ref                 & $0.99 \pm 0.01$ & $0.77 \pm 0.28$ & $0.99 \pm 0.03$ & $0.99 \pm 0.01$ & $0.80 \pm 0.28$ & $0.98 \pm 0.05$ \\
		\midrule
		\multirow{5}{*}{\rotatebox[origin=c]{90}{PubMedQA}} 
		& Loss                & $0.81 \pm 0.07$ & $0.16 \pm 0.11$ & $0.36 \pm 0.15$ & $0.95 \pm 0.03$ & $0.51 \pm 0.21$ & $0.75 \pm 0.14$ \\
		& ZLIB                & $0.77 \pm 0.07$ & $0.10 \pm 0.05$ & $0.30 \pm 0.13$ & $0.88 \pm 0.05$ & $0.32 \pm 0.17$ & $0.57 \pm 0.15$ \\
		& Mink	         	  & $0.86 \pm 0.05$ & $0.25 \pm 0.12$ & $0.48 \pm 0.15$ & $0.98 \pm 0.01$ & $0.75 \pm 0.14$ & $0.91 \pm 0.07$ \\
		& Mink++			  & $0.90 \pm 0.02$ & $0.31 \pm 0.08$ & $0.57 \pm 0.08$ & $0.99 \pm 0.01$ & $0.93 \pm 0.07$ & $0.98 \pm 0.02$ \\
		& Ref                 & $1.00 \pm 0.00$ & $0.98 \pm 0.02$ & $1.00 \pm 0.00$ & $1.00 \pm 0.00$ & $1.00 \pm 0.00$ & $1.00 \pm 0.00$ \\
		\bottomrule
	\end{tabular}
	\label{tab:attack-results}
\end{table}

%% file: DDI_Multi.tex
The DDI results for a two-domain setting appear in
\autoref{tab:ddi-llama-multi} (Llama-3.1-8B) and
\autoref{tab:ddi-mistral-multi} (Mistral-0.1-7B).  As we can see from
these tables, we achieve high DDI values (e.g., close to $\alpha ~=
1.0$ for auc-roc). In other words, an auditor can \emph{almost
perfectly} identify which domain is in effect, even when the
corresponding utility gap ($\Delta_{U}$)
is far below 1.0
(\autoref{fig:utility_gap_simpleqa_llama}). \multiloraunion\ consistently
attains the highest DDI, followed by \multilora\ and then
\multiloramerge\, mirroring the trend observed with $\Delta_{U}$.
\multiloraunion’s superiority however comes at the cost of greater
tuning-time computation. \multiloraunion's near-perfect
distinguishability mirrors the effect of model performance (with
increasing domains) on $\Delta_{U}$ (see
\autoref{fig:utility_gap_simpleqa_llama}).  Crucially, the high DDI
values confirm that even when $\Delta_{U}$ drops due to model
generalization or degradation due to activation space or parameter
interference, access control remains uncompromised; DDI therefore
provides the more sensitive indicator of enforcement efficacy.

% ---------- Llama-3.1-8B only ----------
\begin{table}[tb]
	\centering
	\scriptsize
	\caption{DDI values for models (with base model Llama-3.1-8B)
          with different approaches of combining domains when user has
          access to two domains. All reported values are $mean \pm
          std$ across domains}
	\label{tab:ddi-llama-multi}
	
	\resizebox{\linewidth}{!}{
	\begin{tabular}{ll|ccc|ccc|ccc}
		\toprule
		& MIA & \multicolumn{3}{c}{\multilora}     
		& \multicolumn{3}{c}{\multiloramerge}  
		& \multicolumn{3}{c}{\multiloraunion} \\ 
		& & auc-roc & tpr@1\%fpr & tpr@5\%fpr & auc-roc & tpr@1\%fpr & tpr@5\%fpr & auc-roc & tpr@1\%fpr & tpr@5\%fpr \\
		\midrule
		% ============================  WMDP  ====================================
		\multirow{5}{*}{\rotatebox[origin=c]{90}{WMDP}}
		& Loss      & $0.98 \pm 0.02$ & $0.77 \pm 0.22$ & $0.87 \pm 0.13$ & $0.93 \pm 0.05$ & $0.53 \pm 0.25$ & $0.67 \pm 0.21$ & $0.99 \pm 0.02$ & $0.90 \pm 0.14$ & $0.94 \pm 0.09$ \\
		& ZLIB     & $0.92 \pm 0.08$ & $0.60 \pm 0.27$ & $0.67 \pm 0.28$ & $0.86 \pm 0.09$ & $0.38 \pm 0.21$ & $0.50 \pm 0.26$ & $0.97 \pm 0.05$ & $0.77 \pm 0.31$ & $0.80 \pm 0.28$ \\
		& Mink      & $0.99 \pm 0.01$ & $0.88 \pm 0.08$ & $0.93 \pm 0.04$ & $0.96 \pm 0.02$ & $0.65 \pm 0.19$ & $0.78 \pm 0.12$ & $1.00 \pm 0.00$ & $0.94 \pm 0.08$ & $0.99 \pm 0.01$ \\
		& Mink++    & $0.90 \pm 0.05$ & $0.62 \pm 0.21$ & $0.71 \pm 0.16$ & $0.94 \pm 0.04$ & $0.65 \pm 0.21$ & $0.80 \pm 0.15$ & $1.00 \pm 0.00$ & $1.00 \pm 0.00$ & $1.00 \pm 0.00$ \\
		& Ref       & $1.00 \pm 0.00$ & $0.98 \pm 0.02$ & $0.99 \pm 0.01$ & $0.99 \pm 0.00$ & $0.81 \pm 0.05$ & $0.91 \pm 0.02$ & $1.00 \pm 0.00$ & $0.98 \pm 0.02$ & $1.00 \pm 0.00$ \\
		\midrule
		% ============================  GPQA  ====================================
		\multirow{5}{*}{\rotatebox[origin=c]{90}{GPQA}}
	& Loss      & $0.99 \pm 0.01$ & $0.81 \pm 0.09$ & $0.93 \pm 0.05$ & $0.93 \pm 0.02$ & $0.38 \pm 0.14$ & $0.72 \pm 0.03$  & $1.00 \pm 0.00$ & $0.97 \pm 0.04$ & $0.99 \pm 0.01$ \\
	& ZLIB     & $0.90 \pm 0.06$ & $0.38 \pm 0.26$ & $0.63 \pm 0.22$ & $0.82 \pm 0.07$ & $0.26 \pm 0.17$ & $0.44 \pm 0.16$ & $0.99 \pm 0.01$ & $0.79 \pm 0.30$ & $0.96 \pm 0.05$  \\
	& Mink      & $0.99 \pm 0.01$ & $0.92 \pm 0.11$ & $0.97 \pm 0.04$ & $0.96 \pm 0.01$ & $0.69 \pm 0.07$ & $0.80 \pm 0.07$ & $1.00 \pm 0.00$ & $1.00 \pm 0.00$ & $1.00 \pm 0.00$ \\
	& Mink++   & $0.95 \pm 0.06$ & $0.82 \pm 0.10$ & $0.85 \pm 0.13$  & $0.97 \pm 0.03$ & $0.75 \pm 0.13$ & $0.88 \pm 0.10$ & $1.00 \pm 0.00$ & $1.00 \pm 0.00$ & $1.00 \pm 0.00$ \\
	& Ref       & $1.00 \pm 0.00$ & $0.99 \pm 0.01$ & $0.99 \pm 0.01$ & $0.99 \pm 0.01$ & $0.87 \pm 0.12$ & $0.93 \pm 0.09$ & $1.00 \pm 0.00$ & $1.00 \pm 0.00$ & $1.00 \pm 0.00$ \\
		\midrule
		% ============================  SimpleQA  ===============================
		\multirow{5}{*}{\rotatebox[origin=c]{90}{SimpleQA}}
		& Loss      & $0.96 \pm 0.03$ & $0.42 \pm 0.32$ & $0.73 \pm 0.26$ & $0.95 \pm 0.03$ & $0.47 \pm 0.28$ & $0.74 \pm 0.21$ & $0.97 \pm 0.04$ & $0.62 \pm 0.38$ & $0.83 \pm 0.29$ \\
		& ZLIB     & $0.94 \pm 0.04$ & $0.35 \pm 0.28$ & $0.66 \pm 0.23$ & $0.93 \pm 0.04$ & $0.41 \pm 0.24$ & $0.67 \pm 0.17$ & $0.97 \pm 0.04$ & $0.61 \pm 0.38$ & $0.82 \pm 0.29$ \\
		& Mink      & $0.94 \pm 0.06$ & $0.41 \pm 0.33$ & $0.68 \pm 0.27$ & $0.94 \pm 0.03$ & $0.47 \pm 0.22$ & $0.71 \pm 0.18$ & $0.98 \pm 0.03$ & $0.57 \pm 0.38$ & $0.84 \pm 0.25$ \\
		& Mink++    & $0.85 \pm 0.10$ & $0.25 \pm 0.19$ & $0.57 \pm 0.16$ & $0.92 \pm 0.03$ & $0.34 \pm 0.16$ & $0.62 \pm 0.13$ & $0.97 \pm 0.03$ & $0.57 \pm 0.37$ & $0.85 \pm 0.24$ \\
		& Ref       & $0.96 \pm 0.03$ & $0.37 \pm 0.35$ & $0.73 \pm 0.30$ & $0.96 \pm 0.04$ & $0.43 \pm 0.40$ & $0.69 \pm 0.35$ & $0.97 \pm 0.04$ & $0.58 \pm 0.42$ & $0.79 \pm 0.31$ \\
		\midrule
		% ============================  RCV1  ====================================
		\multirow{5}{*}{\rotatebox[origin=c]{90}{RCV1}}
	& Loss      & $0.96 \pm 0.02$ & $0.40 \pm 0.09$ & $0.76 \pm 0.15$ & $0.90 \pm 0.01$ & $0.24 \pm 0.05$ & $0.52 \pm 0.07$ &$0.98 \pm 0.00$ & $0.55 \pm 0.23$ & $0.94 \pm 0.01$ \\
	& ZLIB     & $0.82 \pm 0.02$ & $0.27 \pm 0.07$ & $0.46 \pm 0.06$ & $0.72 \pm 0.02$ & $0.11 \pm 0.03$ & $0.28 \pm 0.03$ &  $0.90 \pm 0.05$ & $0.52 \pm 0.20$ & $0.67 \pm 0.13$ \\
	& Mink      & $0.97 \pm 0.02$ & $0.60 \pm 0.14$ & $0.87 \pm 0.08$ & $0.92 \pm 0.02$ & $0.29 \pm 0.04$ & $0.65 \pm 0.08$ & $0.99 \pm 0.00$ & $0.80 \pm 0.08$ & $0.97 \pm 0.01$ \\
	& Mink++     & $0.80 \pm 0.13$ & $0.32 \pm 0.19$ & $0.49 \pm 0.24$ & $0.84 \pm 0.07$ & $0.28 \pm 0.22$ & $0.52 \pm 0.19$ & $0.99 \pm 0.00$ & $0.90 \pm 0.05$ & $0.98 \pm 0.00$\\
	& Ref       & $0.97 \pm 0.01$ & $0.50 \pm 0.09$ & $0.86 \pm 0.09$ & $0.95 \pm 0.00$ & $0.26 \pm 0.07$ & $0.63 \pm 0.05$ & $0.98 \pm 0.01$ & $0.50 \pm 0.31$ & $0.95 \pm 0.02$  \\
	\midrule
			\multirow{5}{*}{\rotatebox[origin=c]{90}{PubMedQA}}
	& Loss      & $0.84 \pm 0.05$ & $0.21 \pm 0.11$ & $0.44 \pm 0.12$ & $0.66 \pm 0.02$ & $0.04 \pm 0.01$ & $0.14 \pm 0.01$ &$0.79 \pm 0.04$ & $0.14 \pm 0.05$ & $0.32 \pm 0.09$ \\
	& ZLIB     & $0.78 \pm 0.05$ & $0.13 \pm 0.07$ & $0.33 \pm 0.09$ & $0.61 \pm 0.01$ & $0.03 \pm 0.01$ & $0.12 \pm 0.01$ &$0.72 \pm 0.04$ & $0.09 \pm 0.04$ & $0.25 \pm 0.07$ \\
	& Mink      & $0.91 \pm 0.05$ & $0.39 \pm 0.17$ & $0.63 \pm 0.17$ & $0.73 \pm 0.02$ & $0.07 \pm 0.02$ & $0.22 \pm 0.03$ &$0.87 \pm 0.03$ & $0.25 \pm 0.11$ & $0.49 \pm 0.11$ \\
	& Mink++     & $0.85 \pm 0.10$ & $0.36 \pm 0.22$ & $0.55 \pm 0.23$ & $0.79 \pm 0.05$ & $0.12 \pm 0.05$ & $0.33 \pm 0.08$ &$0.94 \pm 0.02$ & $0.44 \pm 0.17$ & $0.72 \pm 0.12$ \\
	& Ref      & $0.99 \pm 0.02$ & $0.84 \pm 0.16$ & $0.93 \pm 0.08$ & $0.99 \pm 0.01$ & $0.75 \pm 0.18$ & $0.96 \pm 0.04$ &$1.00 \pm 0.00$ & $0.97 \pm 0.07$ & $1.00 \pm 0.00$ \\
		\bottomrule
	\end{tabular}}
\end{table}

% ---------- mistral ----------
\begin{table}[tb]
	\centering
	\scriptsize
	\caption{DDI values for models (with base model
          Mistral-0.1-7B) with different approaches of combining
          domains when user has access to two domains. All reported
          values are $mean \pm std$ across domains.}
	\label{tab:ddi-mistral-multi}
	
	\resizebox{\linewidth}{!}{
	\begin{tabular}{ll|ccc|ccc|ccc}
		\toprule
		& MIA & \multicolumn{3}{c}{\multilora}     
		& \multicolumn{3}{c}{\multiloramerge}  
		& \multicolumn{3}{c}{\multiloraunion} \\ 
		& & auc-roc & tpr@1\%fpr & tpr@5\%fpr & auc-roc & tpr@1\%fpr & tpr@5\%fpr & auc-roc & tpr@1\%fpr & tpr@5\%fpr \\
		\midrule
		% ============================  WMDP  ====================================
		\multirow{5}{*}{\rotatebox[origin=c]{90}{WMDP}}
		& Loss      & $0.99 \pm 0.02$ & $0.85 \pm 0.21$ & $0.92 \pm 0.11$ & $0.95 \pm 0.04$ & $0.62 \pm 0.21$ & $0.73 \pm 0.19$ & $0.99 \pm 0.01$ & $0.93 \pm 0.10$ & $0.96 \pm 0.06$ \\
		& ZLIB     & $0.93 \pm 0.09$ & $0.69 \pm 0.30$ & $0.74 \pm 0.30$ & $0.87 \pm 0.09$ & $0.47 \pm 0.26$ & $0.58 \pm 0.29$ & $0.98 \pm 0.03$ & $0.83 \pm 0.23$ & $0.88 \pm 0.16$ \\
		& Mink      & $0.99 \pm 0.01$ & $0.89 \pm 0.14$ & $0.95 \pm 0.07$ & $0.96 \pm 0.03$ & $0.73 \pm 0.11$ & $0.83 \pm 0.12$ & $1.00 \pm 0.00$ & $1.00 \pm 0.00$ & $1.00 \pm 0.00$ \\
		& Mink++    & $0.96 \pm 0.02$ & $0.77 \pm 0.04$ & $0.86 \pm 0.04$ & $0.94 \pm 0.03$ & $0.58 \pm 0.03$ & $0.80 \pm 0.05$ & $1.00 \pm 0.00$ & $1.00 \pm 0.00$ & $1.00 \pm 0.00$ \\
		& Ref       & $1.00 \pm 0.00$ & $1.00 \pm 0.00$ & $1.00 \pm 0.00$ & $0.99 \pm 0.00$ & $0.86 \pm 0.09$ & $0.96 \pm 0.02$ & $1.00 \pm 0.00$ & $1.00 \pm 0.00$ & $1.00 \pm 0.00$ \\
		\midrule
		% ============================  GPQA  ====================================
		\multirow{5}{*}{\rotatebox[origin=c]{90}{GPQA}}
		& Loss      & $0.99 \pm 0.01$ & $0.83 \pm 0.18$ & $0.95 \pm 0.06$ & $0.96 \pm 0.04$ & $0.55 \pm 0.24$ & $0.87 \pm 0.06$ & $1.00 \pm 0.00$ & $0.97 \pm 0.04$ & $0.98 \pm 0.02$ \\
& ZLIB     & $0.93 \pm 0.08$ & $0.50 \pm 0.35$ & $0.74 \pm 0.32$ & $0.86 \pm 0.09$ & $0.33 \pm 0.23$ & $0.56 \pm 0.21$ & $0.99 \pm 0.01$ & $0.88 \pm 0.17$ & $0.97 \pm 0.04$ \\
& Mink      & $1.00 \pm 0.00$ & $0.94 \pm 0.07$ & $0.98 \pm 0.02$ & $0.98 \pm 0.02$ & $0.74 \pm 0.14$ & $0.87 \pm 0.12$ & $1.00 \pm 0.00$ & $1.00 \pm 0.00$ & $1.00 \pm 0.00$ \\
& Mink++    & $0.98 \pm 0.02$ & $0.80 \pm 0.14$ & $0.92 \pm 0.06$ & $0.98 \pm 0.01$ & $0.75 \pm 0.13$ & $0.89 \pm 0.07$ & $1.00 \pm 0.00$ & $1.00 \pm 0.00$ & $1.00 \pm 0.00$ \\
& Ref       & $1.00 \pm 0.00$ & $1.00 \pm 0.00$ & $1.00 \pm 0.00$ & $0.99 \pm 0.02$ & $0.84 \pm 0.23$ & $0.97 \pm 0.04$ & $1.00 \pm 0.00$ & $0.97 \pm 0.04$ & $1.00 \pm 0.00$ \\
		\midrule
		% ============================  SimpleQA  ===============================
		\multirow{5}{*}{\rotatebox[origin=c]{90}{SimpleQA}}
		& Loss      & $0.97 \pm 0.03$ & $0.58 \pm 0.33$ & $0.82 \pm 0.27$ & $0.96 \pm 0.02$ & $0.49 \pm 0.24$ & $0.79 \pm 0.17$ & $0.97 \pm 0.04$ & $0.50 \pm 0.42$ & $0.76 \pm 0.31$ \\
& ZLIB     & $0.97 \pm 0.03$ & $0.51 \pm 0.32$ & $0.78 \pm 0.28$ & $0.95 \pm 0.03$ & $0.44 \pm 0.23$ & $0.72 \pm 0.19$ & $0.97 \pm 0.04$ & $0.51 \pm 0.42$ & $0.75 \pm 0.31$ \\
& Mink      & $0.97 \pm 0.03$ & $0.51 \pm 0.34$ & $0.83 \pm 0.24$ & $0.96 \pm 0.02$ & $0.49 \pm 0.24$ & $0.77 \pm 0.18$ & $0.97 \pm 0.04$ & $0.51 \pm 0.41$ & $0.79 \pm 0.27$ \\
& Mink++    & $0.92 \pm 0.04$ & $0.46 \pm 0.21$ & $0.68 \pm 0.21$ & $0.93 \pm 0.05$ & $0.45 \pm 0.28$ & $0.73 \pm 0.19$ & $0.97 \pm 0.04$ & $0.50 \pm 0.41$ & $0.76 \pm 0.29$ \\
& Ref       & $0.98 \pm 0.03$ & $0.65 \pm 0.39$ & $0.86 \pm 0.27$ & $0.98 \pm 0.03$ & $0.64 \pm 0.34$ & $0.85 \pm 0.25$ & $0.96 \pm 0.04$ & $0.48 \pm 0.43$ & $0.73 \pm 0.34$ \\
		\midrule
		% ============================  RCV1  ====================================
		\multirow{5}{*}{\rotatebox[origin=c]{90}{RCV1}}
		& Loss      & $0.93 \pm 0.04$ & $0.39 \pm 0.23$ & $0.62 \pm 0.23$ & $0.85 \pm 0.01$ & $0.14 \pm 0.03$ & $0.35 \pm 0.02$ & $0.98 \pm 0.01$ & $0.53 \pm 0.22$ & $0.92 \pm 0.01$ \\
& ZLIB     & $0.82 \pm 0.05$ & $0.30 \pm 0.10$ & $0.50 \pm 0.08$ & $0.69 \pm 0.03$ & $0.10 \pm 0.04$ & $0.26 \pm 0.06$ & $0.90 \pm 0.05$ & $0.48 \pm 0.23$ & $0.67 \pm 0.14$ \\
& Mink      & $0.93 \pm 0.05$ & $0.44 \pm 0.24$ & $0.68 \pm 0.23$ & $0.85 \pm 0.02$ & $0.16 \pm 0.03$ & $0.40 \pm 0.04$ & $0.99 \pm 0.00$ & $0.73 \pm 0.12$ & $0.97 \pm 0.01$ \\
& Mink++    & $0.69 \pm 0.25$ & $0.27 \pm 0.20$ & $0.45 \pm 0.33$ & $0.70 \pm 0.16$ & $0.18 \pm 0.13$ & $0.35 \pm 0.21$ & $0.99 \pm 0.00$ & $0.89 \pm 0.03$ & $0.98 \pm 0.00$ \\
& Ref       & $0.96 \pm 0.02$ & $0.35 \pm 0.12$ & $0.71 \pm 0.18$ & $0.94 \pm 0.01$ & $0.15 \pm 0.05$ & $0.52 \pm 0.08$ & $0.98 \pm 0.00$ & $0.45 \pm 0.25$ & $0.97 \pm 0.00$ \\
\midrule
			\multirow{5}{*}{\rotatebox[origin=c]{90}{PubMedQA}}
& Loss      & $0.78 \pm 0.11$ & $0.19 \pm 0.15$ & $0.38 \pm 0.19$ & $0.65 \pm 0.01$ & $0.04 \pm 0.01$ & $0.13 \pm 0.02$ &$0.81 \pm 0.05$ & $0.16 \pm 0.09$ & $0.36 \pm 0.12$ \\
& ZLIB     & $0.73 \pm 0.10$ & $0.13 \pm 0.10$ & $0.28 \pm 0.14$ & $0.61 \pm 0.01$ & $0.03 \pm 0.01$ & $0.11 \pm 0.01$ &$0.74 \pm 0.05$ & $0.11 \pm 0.04$ & $0.27 \pm 0.18$ \\
& Mink      & $0.83 \pm 0.14$ & $0.30 \pm 0.26$ & $0.49 \pm 0.27$ & $0.70 \pm 0.02$ & $0.05 \pm 0.02$ & $0.18 \pm 0.02$ &$0.89 \pm 0.04$ & $0.26 \pm 0.11$ & $0.53 \pm 0.12$ \\
& Mink++     & $0.72 \pm 0.24$ & $0.29 \pm 0.31$ & $0.44 \pm 0.33$ & $0.74 \pm 0.06$ & $0.08 \pm 0.04$ & $0.23 \pm 0.07$ &$0.95 \pm 0.02$ & $0.53 \pm 0.19$ & $0.76 \pm 0.13$ \\
& Ref      & $0.93 \pm 0.10$ & $0.69 \pm 0.30$ & $0.80 \pm 0.25$ & $0.98 \pm 0.00$ & $0.59 \pm 0.09$ & $0.93 \pm 0.04$ &$1.00 \pm 0.00$ & $0.97 \pm 0.05$ & $1.00 \pm 0.00$ \\
\bottomrule
	\end{tabular}}
\end{table}

%% file: discussion.tex
\section{Conclusion and Discussion}
\label{sec:discussion}

We presented a comprehensive treatment of the access control problem
on fine-tuned LLMs that includes novel formalism, empirical evaluation
metrics, access control enforcement mechanisms, and evaluation of the
mechanisms as well as the proposed metrics.  We formalized a new class
of LLMs called \emph{Permissioned LLMs (\pllm)} whose access control
enforcement can be verified both theoretically and empirically using
the formal tools provided in our work.

{\bf Limitations.} Our approach does not support deep hierarchy of
domains with arbitrary overlaps. Another issue we observe is with the
scalability beyond a handful of domains. This either leads to severe
degradation of utility (as in the case of \multilora) or it becomes
compute-intensive (for \multiloraunion). We leave this exploration for
future work.  We also note some limitations in the experiments that we
do not expect to change our key claims. First, we only run one model
fine-tuning per parameter setting due to the computation
overhead. Second, we use the default value for LoRA rank as our preliminary experiments with different ranks
suggested limited impact on model utility. For our formalism in \autoref{sec:problem_setup}, we assume
that adversaries do not tamper with their credentials or domain
access, otherwise they can gain arbitrary domain information. This is
enforced by the enclosing system via authentication. Finally, we note that our DDI 
metrics are only as good as the best possible MIA.

{\bf Related Work.} Access control problems in agentic systems can
manifest in interesting ways, such as context
hijacking~\citep{bagdasarian24}, and may require further constraining
the purview of individual agent contexts. Retrieval Augmented
Generation (RAG)
systems~\citep{lewis20,owasp25,zhong2025honeybeeefficientrolebasedaccess}
are also susceptible to the access control problem.  However, the
access control needs to be enforced in the information retrieval
engine of the system~\citep{buttcher05,goyal23} and is beyond our
work's scope (although we do provide a formalism for access control in
RAG-based systems in \autoref{sec:rag-formalism}).
One may draw some parallels between our formalism of response
relevance and access advantage metric with prior works on
\emph{indistinguishability}~\citep{arriaga16,bohli20,dwork06,goldwasser88}
in security and privacy.  The mechanisms in this lineage of works are
singularly focused on eliminating distinguishability between the
effects of different data on computations.  In contrast, \pllm's
objective is to maximize domain separation, which implies maximization
of distinguishability.

{\bf Broader Impacts.} 
Our work aims to bolster the security and privacy of individual's data by
enforcing strict access control, such that only people with prior
authorization can get access to the information. Our work is
applicable to healthcare, finance, and more broadly, enterprise / governance applications that
deal with sensitive data of individuals.

%% file: ack.tex
\section{Acknowledgements}

We would like to thank Pallika Kanani and Dan Roth for fruitful discussions that motivated the problem setup of access control in large language model based applications. We would also like to thank David Evans and the anonymous reviewers for providing helpful feedback on the paper.

%% file: appendix.tex
\input{rag-formalism}

\input{algs-proofs}

\section{Audit Games}\label{sec:audit_games}

We formalize black-box games that capture: (i) the distinguishability of security domain-specific responses for DDI, and (ii) the utility disparity induced by access restrictions for UGI. 
Intuitively, in these auditing games, we measure how \emph{effectively} an external auditor can conclude if the access control mechanism is correctly implemented by verifying if the correct domain adapter is activated for a query. This effectiveness is directly correlated with the access advantage score for the target security domain(s). Higher access advantage score denotes \emph{stronger} access control enforcement. A perfectly separated system provides the auditor with an access advantage score of $1.0$.

We consider the same threat setting and auditor privileges for our adversarial games between auditor $\cA$ and system $\cS$ enclosing the \pllm{} $f_D^\cM$ as described in \autoref{sec:threat_model}.

\paragraph{Game 1: Domain Distinguishability.}
This game assesses whether the auditor can effectively conclude
if the correct security domains were used based on the generated
responses. The primary motivation of this game is to measure the
distinguishability of different security domains' distributions.

\begin{enumerate}[itemsep=2pt,   % vertical space between items
	topsep=2pt,    % space before/after the list
	parsep=0pt,    % space between paragraphs in an item
	partopsep=0pt] % extra space when list starts a new part
	\item Auditor $\cA$ chooses security domain set $S_u$ and emulates user $u$. $\cA$ sends user credentials $cred_u$ and query $q \sim \cD_{S_u}$ to system $\cS$. $\cS$ verifies the user credential $cred_u$ and sends back the model response $f_D^\cM(q)$ to $\cA$.
	\item $\cA$ chooses security domain set $S_v$ such that $S_v \cap S_u = \phi$ and emulates user $v$. $\cA$ sends user credentials $cred_v$ and the same query $q \sim \cD_{S_u}$ to $\cS$. $\cS$ verifies the user credential $cred_v$ and sends back the model response $f_D^\cM(q)$ to $\cA$.
	\item $\cA$ sends the models responses and domain information to membership inference oracle $O$ to obtain domain distinguishability score $m(O(f_D^\cM(q)|S_u, f_D^\cM(q)|S_v))$, where $m(\cdot)$ is a membership inference metric (e.g., AUC-ROC or TPR@1\%FPR) in the [0,1] range.
	\item $\cA$ concludes the access control mechanism is correctly implemented if the domain distinguishability score $m(O(f_D^\cM(q)|S_u, f_D^\cM(q)|S_v)) \ge \alpha$.

\end{enumerate}

Note that we can change the above game to distinguish members ($q \sim \cD_{S_u}$) and non-members ($q \sim \cD_{S_v}$) for the target domain set $S_u$, similar to prior MIA setups, which is what we do in our experiments in \autoref{sec:exp}.

\paragraph{Game 2: Utility Gap Evaluation.}
The second game evaluates how distinctly the responses from two
different security domains impact the utility perceived by users. The
rationale behind this game is to confirm that enforced access controls
result in meaningful variations in response quality.

\begin{enumerate}[itemsep=2pt,   % vertical space between items
	topsep=2pt,    % space before/after the list
	parsep=0pt,    % space between paragraphs in an item
	partopsep=0pt] % extra space when list starts a new part
	\item Auditor $\cA$ chooses security domain set $S_u$ and emulates user $u$. $\cA$ sends user credentials $cred_u$ and query $q \sim \cD_{S_u}$ to system $\cS$. $\cS$ verifies the user credential $cred_u$ and sends back the model response $f_D^\cM(q)$ to $\cA$.
	\item $\cA$ chooses security domain set $S_v$ such that $S_v \cap S_u = \phi$ and emulates user $v$. $\cA$ sends user credentials $cred_v$ and the same query $q \sim \cD_{S_u}$ to $\cS$. $\cS$ verifies the user credential $cred_v$ and sends back the model response $f_D^\cM(q)$ to $\cA$.
	\item Given a utility function $U(\cdot)$ (e.g., BLEURT or task accuracy) that outputs values in [0,1] range, $\cA$ concludes the access control mechanism is correctly implemented if the utility gap score $|U(f_D^\cM(q)|S_u) - U(f_D^\cM(q)|S_v)| \ge \alpha$.
\end{enumerate}

We aggregate the utility gaps from this game across all domain set pairs to obtain our UGI metric.

\section{Detailed Experiment Setup}\label{sec:exp_setup}

\subsection{Models}\label{sec:models}
For our instantiation of \pllm, we fine-tune Llama-3.1-8B\citep{llama3} and Mistral-0.1-7B\citep{mistral} pretrained models on four datasets covering multiple distinct security domains (henceforth called \emph{domains}), where we fine-tune a separate LoRA adapter for each domain. To compare our \pllm, we train two additional models with full fine-tuning and LoRA fine-tuning respectively on entire training data. Note that these models are only used for utility baselines as they do not provide access control. For all the LoRA adapters, we use 64 rank and 0.1 dropout. We use AdamW optimizer with 0.1 weight decay to fine-tuned all the models for 5 epochs with 300 warmup steps, 2 batch size and $5 \times 10^{-4}$ learning rate (except for Mistral-0.1-7B full fine-tuning that uses a learning rate of $5 \times 10^{-5}$). We performed grid search over multiple learning rates and warmup steps and found these values to give the best results. For all our experiments, we use 8 H100 GPUs (with 80GB VRAM per GPU), 4 workers per GPU, and 384 GB RAM. One epoch of fine-tuning took from few minutes (for our smallest data set: GPQA) to a couple of hours (for our largest data set: RCV1). Mistral-0.1-7B is released under Apache 2.0 license, and Llama-3.1-8B is released under Llama 3.1 Community License.

\begin{table}[tb]
    \centering \small
    \caption{Data Set Details. Generalization Loss Gap (i.e., gap
      between model's loss on training and test sets) for all models
      are reported after fine-tuning for 5 epochs on each data set.}
    \begin{tabular}{lrrcccccc}
        \toprule
        Data Set & \multicolumn{2}{c}{Data Set Size} & \multicolumn{3}{c}{Llama-3.1-8B Loss Gap} & \multicolumn{3}{c}{Mistral-0.1-7B Loss Gap} \\
        (\# Domains) & Train & Test & Full FT & LoRA & \pllm & Full FT & LoRA & \pllm \\
        \midrule
        WMDP (3)        & 2936  & 732   & 1.96 & 0.52 & 1.15 & 1.36 & 0.65 & 1.07 \\
        GPQA (3)        & 360   & 88    & 2.51 & 1.06 & 1.04 & 1.58 & 0.61 & 1.09 \\
        SimpleQA (10)   & 4089  & 1018  & 2.91 & 0.96 & 1.49 & 1.87 & 0.90 & 1.25 \\
        RCV1 (4)        & 45622 & 22811 & 4.07 & 0.35 & 0.83 & 2.48 & 0.37 & 0.74 \\
        PubMedQA (10)        & 200000 & 11269 & 3.53 & 0.07 & 0.36 & 2.56 & 0.07 & 0.35 \\
        \bottomrule
    \end{tabular}
    \label{tab:datasets}
\end{table}

\subsection{Data Sets}\label{sec:data_sets}
\begin{figure}[tb]
    \centering
    \includegraphics[width=\textwidth]{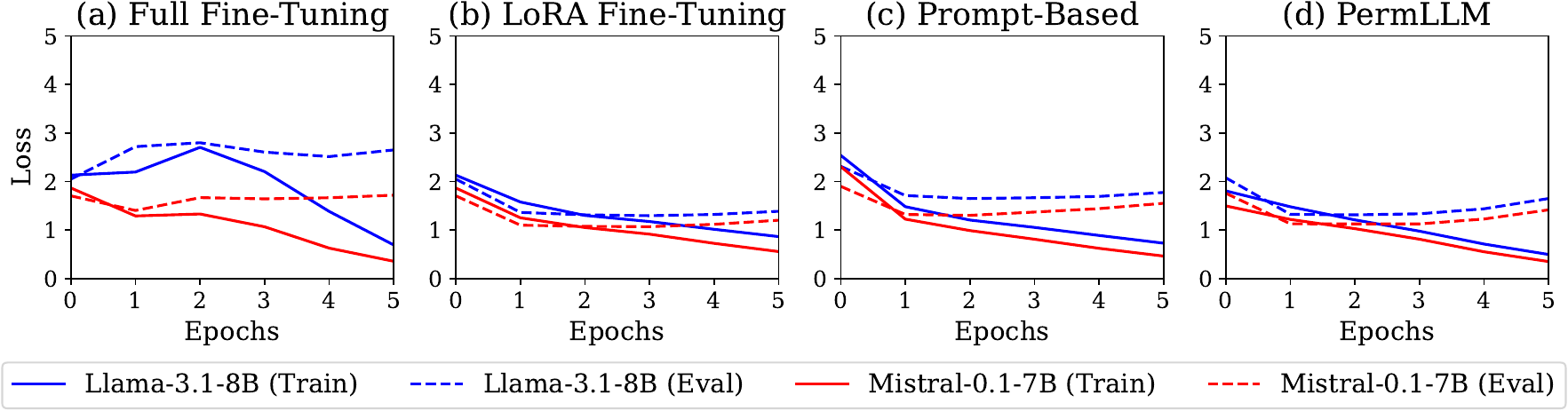}
    \caption{Comparing model loss on WMDP data set.}
    \label{fig:wmdp_loss}
\end{figure}

\begin{figure}[tb]
    \centering
    \includegraphics[width=\textwidth]{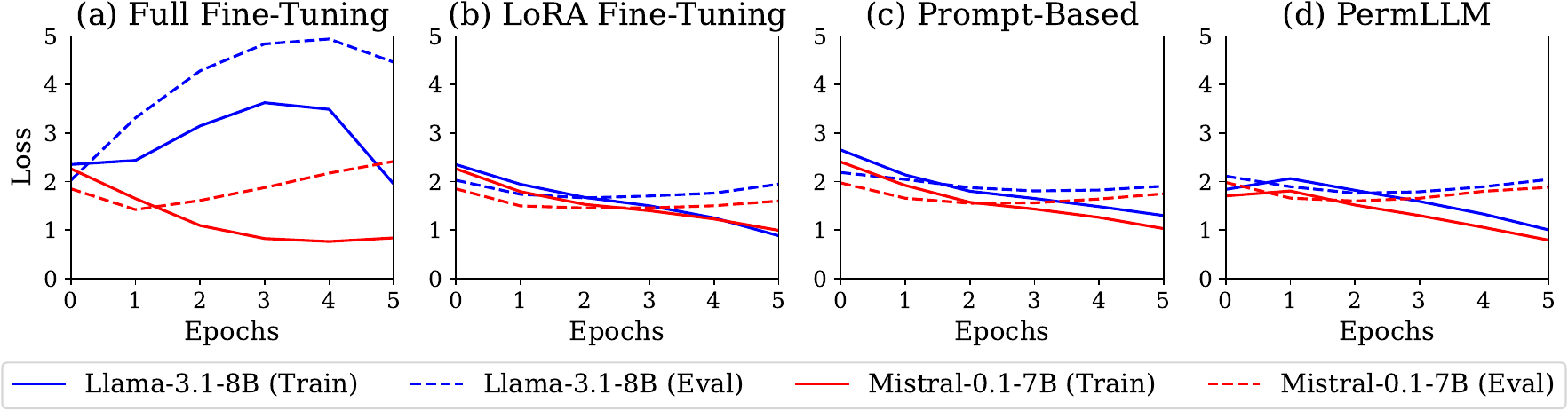}
    \caption{Comparing model loss on GPQA data set.}
    \label{fig:gpqa_loss}
\end{figure}

\begin{figure}[tb]
    \centering
    \includegraphics[width=\textwidth]{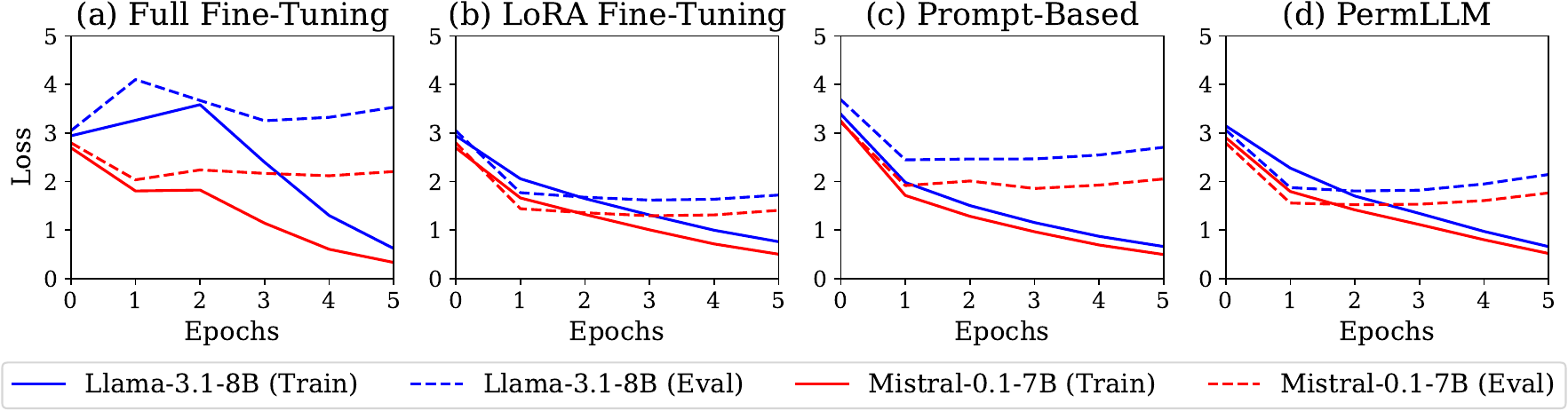}
    \caption{Comparing model loss on SimpleQA data set.}
    \label{fig:simpleqa_loss}
\end{figure}

\begin{figure}[tb]
    \centering
    \includegraphics[width=\textwidth]{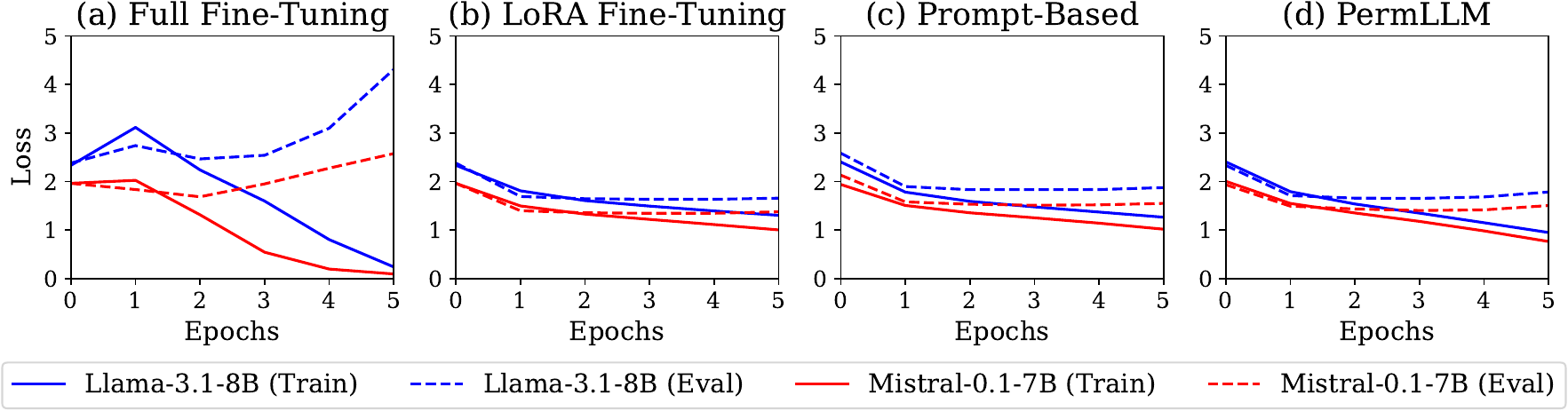}
    \caption{Comparing model loss on RCV1 data set.}
    \label{fig:rcv1_loss}
\end{figure}

\begin{figure}[tb]
    \centering
    \includegraphics[width=\textwidth]{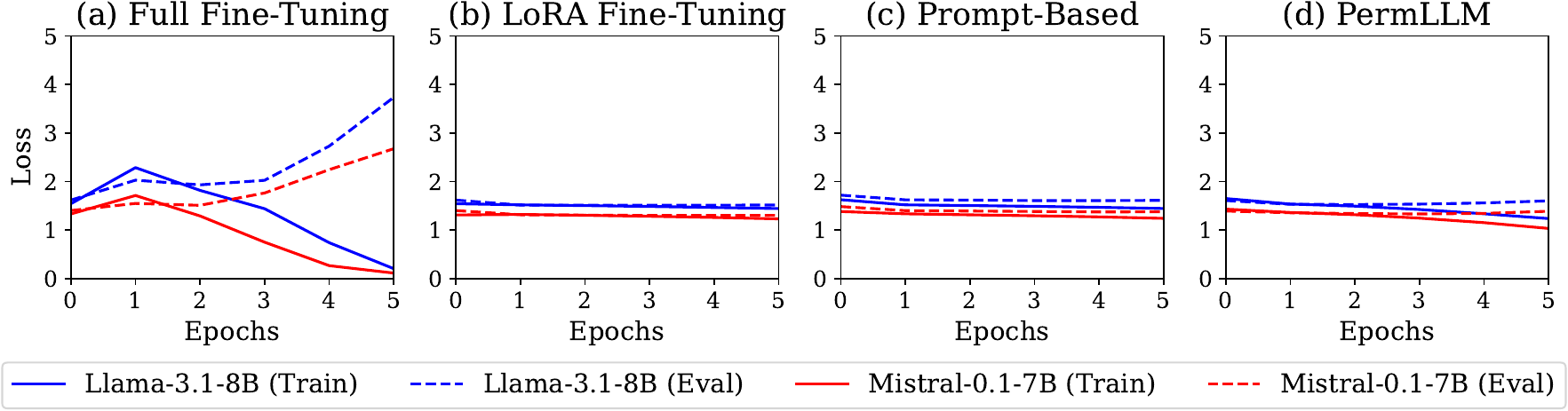}
    \caption{Comparing model loss on PubMedQA data set.}
    \label{fig:pubmedqa_loss}
\end{figure}

For our experiments, we require data sets that
consist of multiple distinct domains and are possibly not seen by the
pretrained models. We use five different data sets, namely,
WMDP~\citep{wmdp}, GPQA~\citep{gpqa}, SimpleQA~\citep{simpleqa},
RCV1~\citep{rcv1}, and PubMedQA~\citep{jin-etal-2019-pubmedqa}. While the first three data sets were collected after
the pretraining cutoff dates for Llama-3.1-8B and Mistral-0.1-7B, RCV1
is an older data set and hence we do not know if it was used in
pretraining. However, we observe a high initial training loss on this
data set, thereby indicating that it was either not used in
pretraining or was catastrophically forgotten by the models, allowing
for a gradual reduction in training loss during our fine-tuning (see
\autoref{fig:rcv1_loss}). \autoref{tab:datasets} shows the
data set details, along with the generalization gap (test loss - train
loss) for different approaches of fine-tuning the models on these data
sets. See \autoref{fig:wmdp_loss}, \autoref{fig:gpqa_loss},
\autoref{fig:simpleqa_loss}, \autoref{fig:rcv1_loss}, and \autoref{fig:pubmedqa_loss}
for complete training and test loss trajectories across different data sets.

\paragraph{WMDP.} Weapons of Mass Destruction Proxy (WMDP)~\citep{wmdp} is a data set consisting of multi-choice question--answer pairs spanning three domains: biological weapons (\emph{bio}), chemical weapons (\emph{chem}) and cyber-warfare weapons (\emph{cyber}). We do 4:1 split of the data set to obtain training and test sets. The training set consists of 2936 question--answer pairs where 1019 are from \emph{bio}, 327 are from \emph{chem} and the remaining 1590 are from \emph{cyber}. The test set size is 732 records, consisting of 254 \emph{bio}, 81 \emph{chem} and 397 \emph{cyber} records. The largest record from this data set consists of 1934 tokens (tokenized using Llama3 tokenizer). This data set is released under MIT License.

\paragraph{GPQA.} Graduate-Level Google-Proof Q\&A Benchmark (GPQA)~\citep{gpqa} data set consists of general question--answer pairs from three domains: \emph{biology}, \emph{chemistry} and \emph{physics}. We do 4:1 split of the data set to obtain training and test sets. The training set consists of 360 question--answer pairs where 63 are from \emph{biology}, 147 are from \emph{chemistry} and the remaining 150 are from \emph{physics}. The test set size is 88 records, consisting of 15 \emph{biology}, 36 \emph{chemistry} and 37 \emph{physics} records. The largest record from this data set consists of 911 tokens (tokenized using Llama3 tokenizer). This data set is released under MIT License.

\paragraph{SimpleQA.} SimpleQA~\citep{simpleqa} is a factuality benchmark that measures the ability for language models to answer short, fact-seeking questions. It consists of general question--answer pairs from ten domains: \emph{art}, \emph{geography}, \emph{history}, \emph{music}, \emph{other}, \emph{politics}, \emph{science and technology}, \emph{sports}, \emph{tv shows}, and \emph{video games}. We do 4:1 split of the data set to obtain training and test sets. The training set consists of 4089 question--answer pairs divided across all ten domains. The test set size is 1018 records spanning across all ten domains. The largest record from this data set consists of 156 tokens (tokenized using Llama3 tokenizer). This data set is released under MIT License.

\paragraph{RCV1.} RCV1~\citep{rcv1} is a benchmark dataset on text categorization. It is a collection of newswire articles produced by Reuters between 1996 and 1997. It contains 804,414 manually labeled newswire documents, broadly categorized with respect to three categories: \emph{industries}, \emph{topics} and \emph{regions}. We took a subset of this data set and created four non-overlapping domains using \emph{topics}: commercial (\emph{CCAT}), economic (\emph{ECAT}), governance (\emph{GCAT}), and mechanical (\emph{MCAT}). We then did 2:1 split of the subset to obtain training and test sets. The training set consists of 45622 question--answer pairs where 23822 are from \emph{CCAT}, 7460 are from \emph{GCAT}, 3370 are from \emph{ECAT} and the remaining 10970 are from \emph{MCAT}. The test set size is 22811 records, consisting of 11911 \emph{CCAT}, 3730 \emph{GCAT}, 1685 \emph{ECAT}, and 5485 \emph{MCAT} records. The largest record from this data set consists of 1199 tokens (tokenized using Llama3 tokenizer). This data set is released under CC BY 4.0 License.

\paragraph{PubMedQA.} PubMedQA~\citep{jin-etal-2019-pubmedqa} contains approximately 200K medical articles formatted as $\langle$Context + Question + Answer$\rangle$. We encoded these articles using the GTE sentence encoder and applied k-means clustering to the resulting embeddings to derive 10 non-overlapping security domains. While clustering enforces semantic similarity within each domain and dissimilarity across domains, the underlying data distribution remains the same, since all samples originate from the same dataset. The largest record from this data set consists of 1614 tokens (tokenized using Llama3 tokenizer). This data set is released under MIT License.

\subsection{Model Utility Evaluation}\label{sec:utility_eval}
We use four metrics to evaluate the utility of the model generations: Bleurt Score ($bluert$), Bert
F1-Score ($bert$), Sacrebleu Score ($bleu$) and Verbatim Accuracy
($acc$). These metrics measure how similar the generated text is to
the ground truth. $bleurt$ and $bert$ measure the semantic similarity,
$bleu$ measures the fraction of common n-grams, and $acc$ gives a
binary decision of whether the generated text verbatim matches the
ground truth. All the metrics lie in a [0,1] range, where values close
to 1 indicate high model utility.

We check the utility of \multilora\ to determine if tuning different
LoRA adapters for each security domain leads to acceptable model
utility. To that end, we show in \autoref{tab:utility_llama} the
utility of Llama-3.1-8B models fine-tuned on different data sets with
the three approaches: full fine-tuning, LoRA fine-tuning and our
\pllm. We do not report the $bleu$ score for WMDP as it is a
multi-choice question-answering task where model only has to generate
a single token. $bleu$ requires generating at least four tokens. Our
approach achieves similar or better utility on the training set
compared to the LoRA approach. On the test set, our approach achieves
similar utility to LoRA for most of the data sets, except for SimpleQA
where LoRA performs better. This is because SimpleQA has more domains
(10 in total), thus each of our individual domain adapter sees only a
fraction of data of what LoRA approach's adapter sees (given that
SimpleQA is already a small data set). We expect the performance of
our domain-specific adapters to increase as the data set size
increases. Full fine-tuning is highly sensitive to training
hyper-parameters, and as a result it either completely overfits on
training set to achieve high utility (e.g., on SimpleQA and RCV1), or
it underfits and achieves low utility (e.g., on WMDP and GPQA). We
observe similar results for Mistral-0.1-7B models (see
\autoref{tab:utility_mistral}).

\begin{table}[tb]
    \centering
    \small
    \caption{Utility comparison of Llama-3.1-8B models trained with different approaches. All reported values are $mean \pm std$ across domains.}
    \begin{tabular}{llcccccc}
        \toprule
        & Metric & \multicolumn{2}{c}{Full Fine-Tuning} & \multicolumn{2}{c}{LoRA Fine-Tuning}  & \multicolumn{2}{c}{\pllm} \\
        & & Train & Test & Train & Test & Train & Test \\
        \midrule
        \multirow{3}{*}{\rotatebox[origin=c]{90}{WMDP}} 
        & $bleurt$    & $0.74 \pm 0.06$ & $0.74 \pm 0.06$ & $0.90 \pm 0.08$ & $0.85 \pm 0.08$ & $0.92 \pm 0.08$ & $0.82 \pm 0.06$ \\
        & $bert$      & $0.89 \pm 0.03$ & $0.89 \pm 0.03$ & $0.96 \pm 0.03$ & $0.94 \pm 0.03$ & $0.97 \pm 0.03$ & $0.93 \pm 0.03$ \\
        & $acc$       & $0.26 \pm 0.07$ & $0.27 \pm 0.07$ & $0.76 \pm 0.20$ & $0.60 \pm 0.20$ & $0.84 \pm 0.22$ & $0.49 \pm 0.15$ \\
        \midrule
        \multirow{4}{*}{\rotatebox[origin=c]{90}{GPQA}} 
        & $bleu$      & $0.26 \pm 0.02$ & $0.05 \pm 0.03$ & $0.45 \pm 0.12$ & $0.10 \pm 0.05$ & $0.39 \pm 0.20$ & $0.10 \pm 0.04$ \\
        & $bleurt$    & $0.53 \pm 0.05$ & $0.39 \pm 0.05$ & $0.64 \pm 0.09$ & $0.46 \pm 0.07$ & $0.62 \pm 0.11$ & $0.47 \pm 0.07$ \\
        & $bert$      & $0.67 \pm 0.06$ & $0.59 \pm 0.05$ & $0.77 \pm 0.08$ & $0.67 \pm 0.05$ & $0.75 \pm 0.09$ & $0.67 \pm 0.05$ \\
        & $acc$       & $0.24 \pm 0.06$ & $0.02 \pm 0.03$ & $0.32 \pm 0.05$ & $0.05 \pm 0.05$ & $0.31 \pm 0.09$ & $0.04 \pm 0.05$ \\
        \midrule
        \multirow{4}{*}{\rotatebox[origin=c]{90}{SimpleQA}} 
        & $bleu$      & $0.80 \pm 0.06$ & $0.34 \pm 0.11$ & $0.65 \pm 0.06$ & $0.29 \pm 0.08$ & $0.67 \pm 0.10$ & $0.09 \pm 0.04$ \\
        & $bleurt$    & $0.86 \pm 0.03$ & $0.58 \pm 0.05$ & $0.80 \pm 0.02$ & $0.61 \pm 0.02$ & $0.82 \pm 0.04$ & $0.53 \pm 0.04$ \\
        & $bert$      & $0.96 \pm 0.01$ & $0.84 \pm 0.02$ & $0.94 \pm 0.01$ & $0.86 \pm 0.01$ & $0.95 \pm 0.02$ & $0.82 \pm 0.03$ \\
        & $acc$       & $0.68 \pm 0.10$ & $0.20 \pm 0.12$ & $0.52 \pm 0.07$ & $0.17 \pm 0.07$ & $0.55 \pm 0.13$ & $0.02 \pm 0.02$ \\
        \midrule
        \multirow{4}{*}{\rotatebox[origin=c]{90}{RCV1}} 
        & $bleu$      & $0.75 \pm 0.08$ & $0.14 \pm 0.08$ & $0.22 \pm 0.10$ & $0.16 \pm 0.08$ & $0.27 \pm 0.10$ & $0.16 \pm 0.08$ \\
        & $bleurt$    & $0.88 \pm 0.04$ & $0.46 \pm 0.12$ & $0.57 \pm 0.13$ & $0.49 \pm 0.11$ & $0.62 \pm 0.13$ & $0.50 \pm 0.12$ \\
        & $bert$      & $0.94 \pm 0.03$ & $0.67 \pm 0.09$ & $0.75 \pm 0.08$ & $0.70 \pm 0.07$ & $0.78 \pm 0.08$ & $0.70 \pm 0.08$ \\
        & $acc$       & $0.78 \pm 0.06$ & $0.16 \pm 0.10$ & $0.27 \pm 0.14$ & $0.17 \pm 0.10$ & $0.31 \pm 0.15$ & $0.18 \pm 0.10$ \\
        \midrule
        \multirow{4}{*}{\rotatebox[origin=c]{90}{\scriptsize PubMedQA}} 
        & $bleu$      & $0.71 \pm 0.05$ & $0.07 \pm 0.01$ & $0.09 \pm 0.01$ & $0.09 \pm 0.01$ & $0.10 \pm 0.02$ & $0.09 \pm 0.01$ \\
        & $bleurt$    & $0.77 \pm 0.03$ & $0.38 \pm 0.01$ & $0.40 \pm 0.01$ & $0.40 \pm 0.01$ & $0.42 \pm 0.01$ & $0.40 \pm 0.02$ \\
        & $bert$      & $0.90 \pm 0.02$ & $0.64 \pm 0.02$ & $0.68 \pm 0.01$ & $0.68 \pm 0.01$ & $0.69 \pm 0.02$ & $0.67 \pm 0.02$ \\
        & $acc$       & - & - & - & - & - & - \\
        \bottomrule
    \end{tabular}
    \label{tab:utility_llama}
\end{table}

\begin{table}[tb]
    \centering
    \small
    \caption{Utility comparison of Mistral-0.1-7B models trained with different approaches. All reported values are $mean \pm std$ across domains.}
    \begin{tabular}{llcccccc}
        \toprule
        & Metric & \multicolumn{2}{c}{Full Fine-Tuning} & \multicolumn{2}{c}{LoRA Fine-Tuning}  & \multicolumn{2}{c}{\pllm} \\
        & & Train & Test & Train & Test & Train & Test \\
        \midrule
        \multirow{3}{*}{\rotatebox[origin=c]{90}{WMDP}} 
        & $bleurt$    & $0.95 \pm 0.01$ & $0.82 \pm 0.03$ & $0.96 \pm 0.02$ & $0.87 \pm 0.03$ & $0.96 \pm 0.01$ & $0.86 \pm 0.03$ \\
        & $bert$      & $0.98 \pm 0.01$ & $0.92 \pm 0.02$ & $0.99 \pm 0.01$ & $0.94 \pm 0.02$ & $0.99 \pm 0.01$ & $0.94 \pm 0.02$ \\
        & $acc$       & $0.88 \pm 0.04$ & $0.46 \pm 0.14$ & $0.92 \pm 0.07$ & $0.60 \pm 0.09$ & $0.93 \pm 0.04$ & $0.58 \pm 0.11$ \\
        \midrule
        \multirow{4}{*}{\rotatebox[origin=c]{90}{GPQA}} 
        & $bleu$      & $0.46 \pm 0.03$ & $0.06 \pm 0.05$ & $0.35 \pm 0.08$ & $0.11 \pm 0.07$ & $0.55 \pm 0.18$ & $0.13 \pm 0.06$ \\
        & $bleurt$    & $0.65 \pm 0.04$ & $0.42 \pm 0.08$ & $0.59 \pm 0.09$ & $0.47 \pm 0.06$ & $0.67 \pm 0.09$ & $0.47 \pm 0.08$ \\
        & $bert$      & $0.75 \pm 0.05$ & $0.62 \pm 0.07$ & $0.73 \pm 0.08$ & $0.68 \pm 0.05$ & $0.79 \pm 0.08$ & $0.66 \pm 0.09$ \\
        & $acc$       & $0.38 \pm 0.04$ & $0.04 \pm 0.05$ & $0.24 \pm 0.04$ & $0.05 \pm 0.06$ & $0.40 \pm 0.09$ & $0.08 \pm 0.02$ \\
        \midrule
        \multirow{4}{*}{\rotatebox[origin=c]{90}{SimpleQA}} 
        & $bleu$      & $0.94 \pm 0.02$ & $0.36 \pm 0.11$ & $0.73 \pm 0.06$ & $0.34 \pm 0.09$ & $0.70 \pm 0.13$ & $0.10 \pm 0.04$ \\
        & $bleurt$    & $0.94 \pm 0.01$ & $0.60 \pm 0.04$ & $0.84 \pm 0.03$ & $0.62 \pm 0.03$ & $0.83 \pm 0.06$ & $0.52 \pm 0.04$ \\
        & $bert$      & $0.99 \pm 0.01$ & $0.85 \pm 0.02$ & $0.96 \pm 0.01$ & $0.87 \pm 0.01$ & $0.95 \pm 0.03$ & $0.82 \pm 0.03$ \\
        & $acc$       & $0.91 \pm 0.04$ & $0.23 \pm 0.12$ & $0.62 \pm 0.08$ & $0.20 \pm 0.10$ & $0.60 \pm 0.16$ & $0.03 \pm 0.02$ \\
        \midrule
        \multirow{4}{*}{\rotatebox[origin=c]{90}{RCV1}} 
        & $bleu$      & $0.92 \pm 0.06$ & $0.17 \pm 0.09$ & $0.28 \pm 0.13$ & $0.20 \pm 0.10$ & $0.37 \pm 0.14$ & $0.19 \pm 0.09$ \\
        & $bleurt$    & $0.93 \pm 0.02$ & $0.48 \pm 0.12$ & $0.60 \pm 0.13$ & $0.51 \pm 0.12$ & $0.66 \pm 0.12$ & $0.50 \pm 0.12$ \\
        & $bert$      & $0.98 \pm 0.02$ & $0.69 \pm 0.08$ & $0.78 \pm 0.09$ & $0.71 \pm 0.08$ & $0.81 \pm 0.08$ & $0.71 \pm 0.08$ \\
        & $cc$       & $0.92 \pm 0.03$ & $0.19 \pm 0.11$ & $0.31 \pm 0.15$ & $0.20 \pm 0.11$ & $0.38 \pm 0.17$ & $0.19 \pm 0.10$ \\
        \midrule
        \multirow{4}{*}{\rotatebox[origin=c]{90}{\scriptsize PubMedQA}} 
        & $bleu$      & $0.75 \pm 0.04$ & $0.08 \pm 0.01$ & $0.09 \pm 0.01$ & $0.08 \pm 0.01$ & $0.11 \pm 0.02$ & $0.08 \pm 0.01$ \\
        & $bleurt$    & $0.80 \pm 0.03$ & $0.39 \pm 0.01$ & $0.41 \pm 0.01$ & $0.41 \pm 0.01$ & $0.43 \pm 0.02$ & $0.41 \pm 0.01$ \\
        & $bert$      & $0.92 \pm 0.01$ & $0.65 \pm 0.02$ & $0.69 \pm 0.01$ & $0.68 \pm 0.02$ & $0.70 \pm 0.02$ & $0.68 \pm 0.01$ \\
        & $acc$       & - & - & - & - & - & - \\
        \bottomrule
    \end{tabular}
    \label{tab:utility_mistral}
\end{table}

\begin{figure}[tb]
    \includegraphics[width=\textwidth]{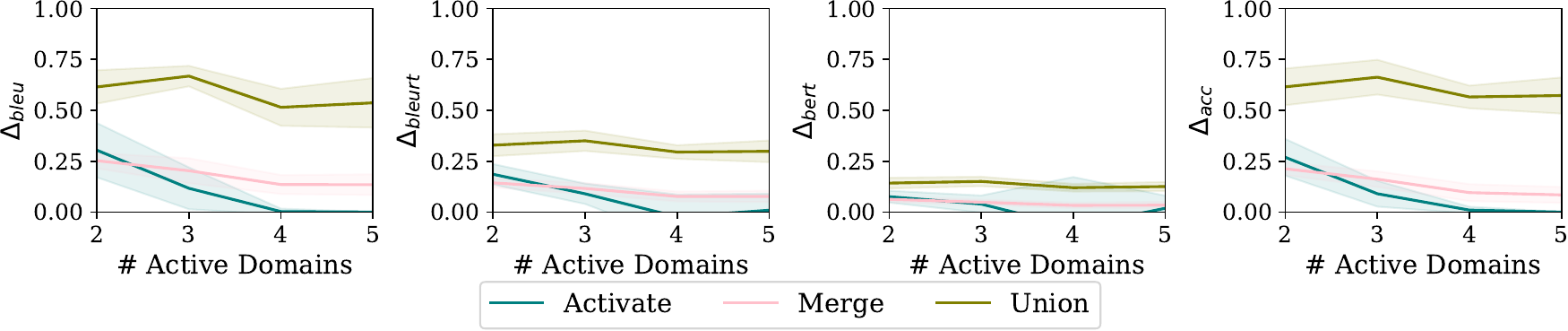}
    \caption{Utility Gap Index, $\Delta_U$ ($mean \pm std$) for Mistral-0.1-7B models fine-tuned on SimpleQA when user has access to multiple security domains.}
    \label{fig:utility_gap_simpleqa_mistral}
\end{figure}

\section{Prompt-Based Access Control}\label{sec:prompt_based}

\begin{figure}[tb]
    \centering
    \includegraphics[width=\textwidth]{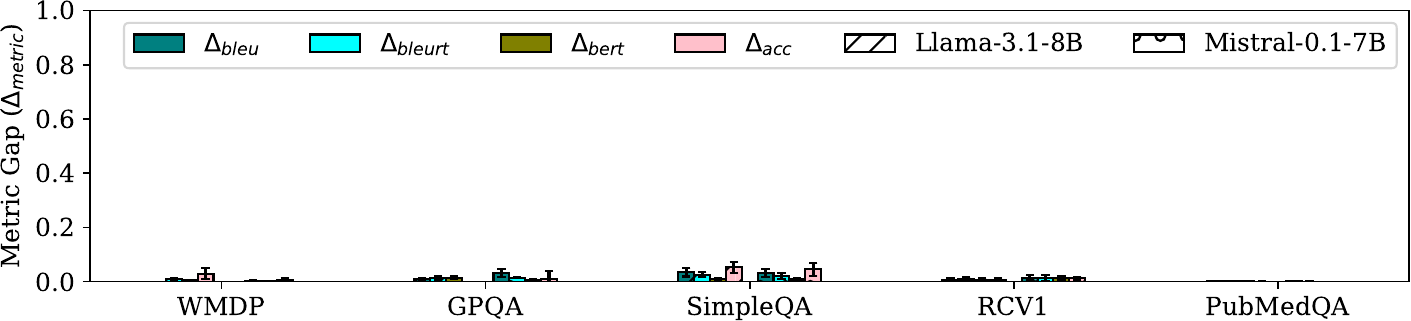}
    \caption{Utility Gap Index, $\Delta_U$ ($mean \pm std$) for prompt-based access control baseline when user has access to one security domain.}
    \label{fig:utility_Gap_prompt_based}
\end{figure}

\begin{figure}[tb]
    \centering
    \includegraphics[width=\textwidth]{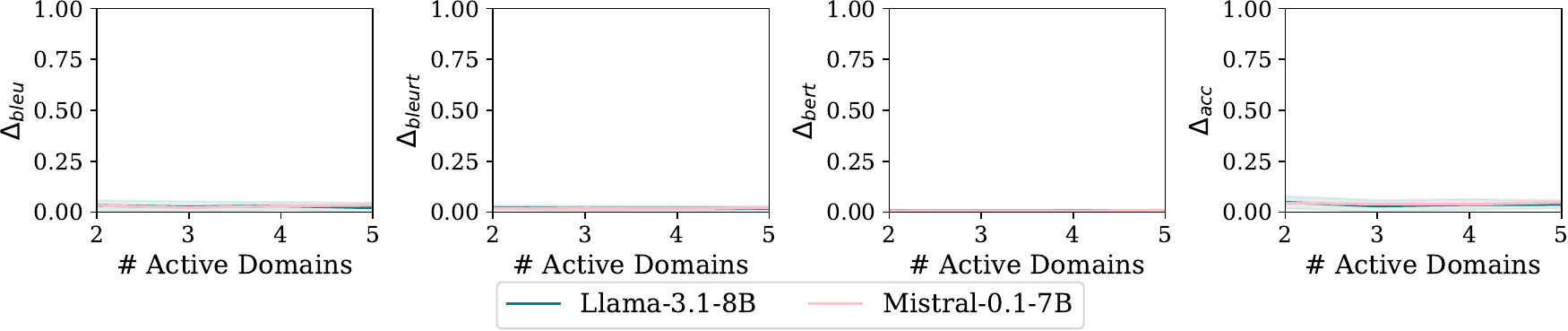}
    \caption{Utility Gap Index, $\Delta_U$ ($mean \pm std$) for prompt-based access control baseline on different models fine-tuned on SimpleQA when user has access to multiple security domains.}
    \label{fig:utility_gap_simpleqa_prompt_based}
\end{figure}

Recent works~\citep{chen23a,liu25} have proposed enforcing some form of access control in system prompts, however we note that they do not provide absolute access control and are vulnerable to jailbreaking prompts. Regardless, we implement prompt-based access control as a baseline where each query is tagged with a prompt prefix (e.g., ``\emph{use domain 1}'') and the rest of the fine-tuning pipeline is similar to LoRA fine-tuning. We add the relevant prompt prefixes during both model fine-tuning and inference. The models fine-tuned with prompt-based access control achieve similar training and test loss to that of LoRA fine-tuning across all the data sets, as shown in \autoref{fig:wmdp_loss}, \autoref{fig:gpqa_loss},
\autoref{fig:simpleqa_loss}, \autoref{fig:rcv1_loss}, and \autoref{fig:pubmedqa_loss}. However, this baseline fails to provide any meaningful access control, even when a user has access to only one security domain as shown in \autoref{fig:utility_Gap_prompt_based} and \autoref{tab:ddi-promptbased_onesecuritydomain}. As shown in the figure and table, the utility gap index is close to \emph{zero} and DDI scores are close to random guessing across all the data sets for both Llama and Mistral models fine-tuned with prompt-based access control. The reason is that the prompt prefix for different domains only differ in one or two tokens and hence the model tends to ignore this difference and continues generating responses even for domains the user has no access to. Exploring different prompt structures might lead to better access control but is beyond the scope of this work. We observe a similar trend when the user has access to multiple security domains as shown in \autoref{fig:utility_gap_simpleqa_prompt_based} for SimpleQA data set.

% ---------- Llama-3.1-8B only ----------
\begin{table}[tb]
	\centering
	\scriptsize
	\caption{DDI values for prompt-based access control baseline when user has access to one security domain. }
	\label{tab:ddi-promptbased_onesecuritydomain}
	
	\resizebox{\linewidth}{!}{
		\begin{tabular}{ll|ccc|ccc}
			\toprule
			& MIA & \multicolumn{3}{c}{Llama-3.1-8B}     
			& \multicolumn{3}{c}{Mistral-0.1-7B}  \\
			& & auc-roc & tpr@1\%fpr & tpr@5\%fpr & auc-roc & tpr@1\%fpr & tpr@5\%fpr \\
			\midrule
			% ============================  WMDP  ====================================
			\multirow{5}{*}{\rotatebox[origin=c]{90}{WMDP}}
			& Loss      & $0.53 \pm 0.02$ & $ 0.02 \pm 0.01 $ & $ 0.06 \pm 0.01$ & $0.54 \pm 0.02$ & $0.02 \pm 0.01$ & $0.07 \pm 0.02$  \\
			& ZLIB     & $0.52 \pm 0.01$ & $0.01 \pm 0.01$ & $0.06 \pm 0.01$ & $0.52 \pm 0.01$ & $0.01 \pm 0.01$ & $0.06 \pm 0.01$  \\
			& Mink      & $ 0.53 \pm 0.03$ & $0.02 \pm 0.02$ & $0.08 \pm 0.03$ & $0.53 \pm 0.01$ & $0.02 \pm 0.01$ & $0.06 \pm 0.01$  \\
			& Mink++    & $0.55 \pm 0.06$ & $0.02 \pm 0.03$ & $0.08 \pm 0.05$ & $0.52 \pm 0.03$ & $0.01 \pm 0.00$ & $0.05 \pm 0.01$  \\
			& Ref       & $0.53 \pm 0.02$ & $0.02 \pm 0.01$ & $0.06 \pm 0.00$ & $0.53 \pm 0.01$ & $0.01 \pm 0.01$ & $0.06 \pm 0.01$  \\
			\midrule
			% ============================  GPQA  ====================================
			\multirow{5}{*}{\rotatebox[origin=c]{90}{GPQA}}
			& Loss      & $0.55 \pm 0.02$ & $0.02 \pm 0.00$ & $0.06 \pm 0.10$ & $0.56 \pm 0.03$ & $0.03 \pm 0.01$ & $0.12 \pm 0.04$  \\
			& ZLIB     & $0.54 \pm 0.02$ & $0.02 \pm 0.00$ & $0.07 \pm 0.01$ & $0.54 \pm 0.02$ & $0.03 \pm 0.01$ & $0.08 \pm 0.02$  \\
			& Mink      & $0.57 \pm 0.05$ & $0.02 \pm 0.01$ & $0.12 \pm 0.02$ & $0.59 \pm 0.07$ & $0.05 \pm 0.06$ & $0.13 \pm 0.07$  \\
			& Mink++    & $0.54 \pm 0.10$ & $ 0.05 \pm 0.06$ & $0.12 \pm 0.08$ & $0.55 \pm 0.12$ & $0.06 \pm 0.06$ & $0.12 \pm 0.09$  \\
			& Ref       & $0.57 \pm 0.02$ & $0.04 \pm 0.02$ & $0.13 \pm 0.05$ & $0.56 \pm 0.03$ & $0.03 \pm 0.02$ & $0.13 \pm 0.07$  \\
			\midrule
			% ============================  SimpleQA  ===============================
			\multirow{5}{*}{\rotatebox[origin=c]{90}{SimpleQA}}
			& Loss      & $0.53 \pm 0.25$ & $0.08 \pm 0.14$ & $0.16 \pm 0.20$ & $0.55 \pm 0.22$ & $0.09 \pm 0.15$ & $0.16 \pm 0.20$  \\
			& ZLIB     & $0.52 \pm 0.16$ & $0.04 \pm 0.05$ & $0.09 \pm 0.09$ & $0.53 \pm 0.14$ & $0.03 \pm 0.03$ & $0.09 \pm 0.08$  \\
			& Mink      & $0.52 \pm 0.28$ & $0.09 \pm 0.15$ & $0.17 \pm 0.22$ & $0.55 \pm 0.22$ & $0.09 \pm 0.15$ & $0.17 \pm 0.20$  \\
			& Mink++    & $0.50 \pm 0.43$ & $0.31 \pm 0.40$ & $0.36 \pm 0.43$ & $0.52 \pm 0.35$ & $0.22 \pm 0.33$ & $0.28 \pm 0.35$  \\
			& Ref       & $0.53 \pm 0.22$ & $0.03 \pm 0.03$ & $0.11 \pm 0.12$ & $0.54 \pm 0.15$ & $0.04 \pm 0.06$ & $0.09 \pm 0.09$  \\
			\midrule
			% ============================  RCV1  ====================================
			\multirow{5}{*}{\rotatebox[origin=c]{90}{RCV1}}
			& Loss      & $0.50 \pm 0.02$ & $0.01 \pm 0.00$ & $0.05 \pm 0.01$ & $0.50 \pm 0.01$ & $0.01 \pm 0.00$ & $0.05 \pm 0.00$  \\
			& ZLIB     & $0.50 \pm 0.01$ & $0.01 \pm 0.00$ & $0.05 \pm 0.02$ & $0.50 \pm 0.00$ & $0.01 \pm 0.00$ & $0.05 \pm 0.00$  \\
			& Mink      & $0.50 \pm 0.04$ & $0.01 \pm 0.00$ & $0.05 \pm 0.01$ & $0.50 \pm 0.01$ & $0.01 \pm 0.02$ & $0.05 \pm 0.01$  \\
			& Mink++    & $0.50 \pm 0.05$ & $0.01 \pm 0.01$ & $0.05 \pm 0.01$ & $0.50 \pm 0.04$ & $0.01 \pm 0.00$ & $0.05 \pm 0.01$  \\
			& Ref       & $0.50\pm 0.01$ & $0.01 \pm 0.01$ & $0.05 \pm 0.01$ & $0.50 \pm 0.01$ & $0.01 \pm 0.00$ & $00.5 \pm 0.00$  \\
			\midrule
			
			\multirow{5}{*}{\rotatebox[origin=c]{90}{PubMedQA}}
			& Loss      & $0.50 \pm 0.00$ & $0.01 \pm 0.00$ & $0.05 \pm 0.00$ & $0.50 \pm 0.00$ & $0.01 \pm 0.00$ & $0.05 \pm 0.00$  \\
			& ZLIB      & $0.50 \pm 0.00$ & $0.01 \pm 0.00$ & $0.05 \pm 0.00$ & $0.50 \pm 0.00$ & $0.01 \pm 0.00$ & $0.05 \pm 0.00$  \\
			& Mink      & $0.50 \pm 0.01$ & $0.01 \pm 0.00$ & $0.05 \pm 0.00$ & $0.50 \pm 0.01$ & $0.01 \pm 0.00$ & $0.05 \pm 0.01$  \\
			& Mink++     & $0.50 \pm 0.01$ & $0.01 \pm 0.00$ & $0.05 \pm 0.00$ & $0.50 \pm 0.01$ & $0.01 \pm 0.00$ & $0.05 \pm 0.00$ \\
			& Ref        & $0.50 \pm 0.03$ & $0.01 \pm 0.00$ & $0.05 \pm 0.01$ & $0.50 \pm 0.03$ & $0.01 \pm 0.00$ & $0.05 \pm 0.01$  \\
			\bottomrule
	\end{tabular}}
\end{table}

\input{MIAs_app}

%% file: rag-formalism.tex
\section{Formalizing Access Control for Retrieval Augmented Generation}
\label{sec:rag-formalism}

For Retrieval Augmented Generation (RAG), we assume a pre-trained LLM
$f$ that is used in applications without additional fine-tuning.
Instead, we augment $f$ with a \emph{retriever engine} $R$ to give us
a retrieval augmented LLM $f_R$.

Each query $q_c$ to $f_R$ is accompanied by a context $c$, retrieved
by $R$, that enhances $f_R$'s response to the query.  Let $R$ retrieve
contexts from the context database $C$, i.e.~$c \in C$.  Furthermore,
we have $C = \bigcup_{i=1}^n C_{s_i} \sim \cC_{s_i}$, where each
$C_{s_i}$ is a collection of contexts belonging to security domain
$s_i$.

For this discussion, we define $\cM$ as an access control mechanism
that dictates the mapping of every $C_{s_i} \subseteq C$ to the
security domain $s_i$.  We say that a RAG system that uses contexts
from the context database $C$ is \emph{permissioned} (\prag), if it
uses retriever $R_C^\cM$, which in turn uses the access control
mechanism $\cM$ to retrieve context $c \in C_{s_i}$ from a selected
security domain $s_i$.  Intuitively, given a security domain $s_i$,
$R$ uses $\cM$ to retrieve context $c \in C_{s_i}$.  One can trivially
generalize this definition of \prag\ to work with subsets of security
domains instead of a singleton security domain $s_i$.

For \prag, we assume an identical enclosing system setting as in
\pllm\ (see~\autoref{sec:problem_setup}): Given a user $u$ the
enclosing system determines $u$'s access credentials $cred_u$ and
calls \texttt{authenticate($cred_u$)} that takes user credentials
$cred_u$ and maps them to a subset of security domains $S_u$ that $u$
can access.  User $u$ cannot arbitrarily change $S_u$.  Each of user
$u$'s subsequent query $q$ to $f_R$ is annotated with $S_u$.  The
retriever $R_C^\cM$ of $f_R$ uses access control mechanism $\cM$ to
retrieve a context $c \in C_{S_u}$.

\begin{definition}[Relevant Response for \prag]
  Given a \prag\ $f_R$, with retriever $R_C^\cM$, a query $q$ from
  user $u$, and $S_u$ the security domains $u$ has access to, $r =
  f_R(q)$ is the response by $f_R$ to query $q$. Response $r$ is
  said to be relevant to $S_u$ (i.e.~$r = r_{S_u}$) if retriever
  $R_C^\cM$ uses a context $c \in C_{S_u}$ to augment the query for
  $r$.
\label{def:relevant-response-prag}
\end{definition}

To empirically quantify response relevance, we can use the same
response relevance score, $relv(f_R(q), S_u)$ that quantifies the
information gained on data in the security domains $q$'s user $u$ has
access to (this is the same set of security domains that mapping $\cM$
gives for $u$ for the retriever $R_C^\cM$, i.e.~$S_u$).  Here
$R_C^\cM$ retrieves the query context $c \in C$ using mapping $\cM$;
$c$ is then augmented to the query $q$.  We restrict the domain of
\relv\ to the real number interval $[0,1]$, where $1$ is the best
expected score for relevance.  Similar to \pllm, we define
\emph{access advantage} for \prag\ as follows:

\begin{definition}[Access Advantage for \prag]
  Given \prag\ $f_R$ that uses retriever $R_C^\cM$ which in turn uses
  the context database $C$ containing data from domains $\bS =
  \bigcup_{i=1}^n \{s_i\}$, with access control mechanism $\cM$, a subset
  of security domains $S_u \subseteq \bS$, context $c \in C$ that is
  augmented to query $q$, $f_R$ achieves \emph{$\alpha$-access
  advantage} w.r.t. $S_u$ if:
  \begin{equation*}
    \bE_{q \sim \cD_{S_u}, S_v \subseteq
    \bS; S_u \cap S_v = \phi}
    \big[ relv(f_R(q), S_u) \circleddash
      relv(f_R(q), S_v) \big] \ge \alpha
  \end{equation*}
  where $relv()$ is the response relevance score on the corresponding
  security domain subset ($S_u$ or $S_v$), $\circleddash$ is a
  ``difference'' operator specific to the access control assessment
  method (e.g. subtraction), and $\alpha$ is an advantage threshold
  that lies in the range [0,1].
  \label{def:access-adv-prag}
\end{definition}

%% file: algs-proofs.tex
\section{Formal Access Control Enforcement in \pllm\ Mechanisms}
\label{sec:algs-formalism}

We now present formal proofs for correct access control enforcement in
our \pllm\ mechanisms presented in~\autoref{sec:algos}: \multilora,
\multiloramerge, and \multiloraunion.

Refreshing the formalism from~\autoref{sec:problem_setup}, we consider
a universe of $n$ different security domains $\bS = \bigcup_{i=1}^n
\{s_i\}$, and a training data set consisting of data from these domains $D
= \bigcup_{i = 1}^n D_{s_i} \sim \cD_{s_i}$ (here $D_{s_i}$ is a data
set sampled from data distribution $\cD_{s_i}$ of domain $s_i$).  Let
$f_{D}$ be the LLM tuned using data set $D$.  Let $W$ be the set of
$f_{D}$'s parameters.  Model tuning \emph{changes} values of a subset
of $W$.  Let security domain $s_i$ \emph{affect}, per the meaning of
affect in~\autoref{sec:problem_setup}, a subset of parameters $W_{s_i}
\subseteq W$.  Thus data from $D_{s_i}$ is used to change parameters
$W_{s_i}$ during model fine-tuning.  Let $\cM$ be the access control
mechanism that dictates the mapping of security domain $s_i$ to
parameters $W_{s_i}$ via the affects relation.

Consider a set of LoRA adapters~\citep{hu21} $l_1,l_2,...,l_m$.  Each
adapter $l_i$ comprises parameters $W_{l_i}$, such that $W_{l_i} \cap
W_{l_j} = \phi, \forall i \neq j$.  Let $i$ be the adapter Id for
adapter $l_i$.  Let $f_D^\cM$ by the \pllm\ that uses mapping $\cM$ of
security domains to parameters during tuning and testing.  Let
$\cF^\cM$ be the system enclosing $f_D^\cM$ that performs the mapping
from user credentials $cred_u$ to sets of security domains $S_u$ for
each user $u$.  We make two assumptions about $\cF^\cM$: (i) $\cF^\cM$
can correctly determine and maintain the security domains $S_u$ a user
$u$ has access to; and (ii) $S_u$ remains opaque to the user and any
other adversary and as a result, cannot be tampered with by any user
or adversary.

We assume that both fine-tuning and testing are mediated through
$\cF^\cM$.  During fine-tuning, the dataset $D$ is passed to $\cF^\cM$.
$\cF^\cM$ extracts information about the security domains $s_1,..,s_n$
covered by $D$.  For settings where users have access to multiple
security domains, the list of security domain combinations that users
have access to is also passed on to $\cF^\cM$.  $\cF^\cM$ does the
mapping between security domain groups and LoRA adapters differently
for each of our \pllm\ mechanisms:
\begin{itemize}
  \item[\multilora] $\cF^\cM$ maps each security domain $s_i$ to a
    unique LoRA adapter $l_i$.  For fine-tuning of $f_D^\cM$,
    minibatches sampled for each $s_i$ are routed to the corresponding
    LoRA adapter $l_i$, the other LoRA adapters are deactivated for
    the sampled mini-batch.
  \item[\multiloramerge] Security domain-LoRA adapter mappings and
    fine-tuning of $f_D^\cM$ proceeds identically to that in
    \multilora.  However, after the fine-tuning is done, the security
    domain groups are used to merged LoRA adapters.  These new LoRA
    adapters are added to the set of LoRA adapters in $f_D^\cM$. The
    mapping $\cM$ is also updated with the new mappings between
    security domain groups and LoRA adapters.
  \item[\multiloraunion] Datasets corresponding to the security domain
    groups are used to fine-tuning unique LoRA adapters.  $\cM$ is
    also updated with these new security domain group-LoRA adapter
    mappings.
\end{itemize}

At the end of fine-tuning, $\cM$ will have a mapping between each
security domain group $S_u$ (for each respective user $u$) and each
LoRA adapter in mechanisms \multiloramerge\ and \multiloraunion.  More
formally,

\begin{lemma}
  In \multiloramerge\ and \multiloraunion, after fine-tuning, for
  every user $u$ that has access to $S_u \subseteq \bS, \exists
  l_{S_u}$, where $l_{S_u}$ is a LoRA adapter, $S_u$ \emph{affects}
  parameters $W_{l_{S_u}}$, and $W_{l_{S_u}}$ is \emph{not} affected
  by any other security domains in $S$.
\label{lemma:affect-multilora-merge-union}
\end{lemma}

In case of \multilora, $S_u$ is used at inference time to activate the LoRA
adapters $l_{s_i}$, where $s_i \in S_u$.  More formally,

\begin{lemma}
  In \multilora, after fine-tuning, for every user $u$ that has access
  to $S_u \subseteq \bS$, $\forall s_i \in S_u$, $s_i$ \emph{affects}
  parameters $W_{l_{s_i}}$, and $W_{l_{s_i}}$ is \emph{not} affected by
  any other security domain $s_j \in S_u, i \neq j$, or $s_k \in \bS \setminus
  S_u$.
\label{lemma:affect-multilora}
\end{lemma}

At inference time, user $u$ sends query $q$ to $\cF^\cM$.  $\cF^\cM$ first
determines $u$'s security domains $S_u$, and then passes $q$ and $S_u$
to $f_D^\cM$, which then activates the LoRA adapter/s corresponding to
$S_u$: $l_{S_u}$ in case of \multiloramerge\ and \multiloraunion, and
$l_{s_i}$, where $s_i \in S_u$, in case of \multilora.  Our
assumptions about accessibility of $S_u$ to the user or adversary
ensure that the adversary cannot tamper with $S_u$ within the scope of
$\cF^\cM$.

\begin{theorem}
  Given any query $q$ from any user $u$, the response $r = f_D^\cM(q)$
  is relevant to $S_u$ for $\cM$ in $\multilora$, $\multiloramerge$,
  or $\multiloraunion$.
\label{thm:multilora-merge-union}
\end{theorem}
\begin{proof}
  From Lemmas~\autoref{lemma:affect-multilora-merge-union}
  and~\autoref{lemma:affect-multilora}, through the fine-tuning
  process $S_u$ affects parameters $W_{l_{S_u}}$ in
  \multiloramerge\ and \multiloraunion, and parameters $W_{l_{s_i}},
  \forall s_i \in S_u$ in \multilora.  At inference time, these same
  parameters (along with the pretrained model's parameters) are used
  to generate response $r = f_D^\cM(q)$.  By implication, the
  parameters affected by $S_u$ are used to generated $r$.  Hence $r$
  is relevant to $S_u$, i.e. $r = r_{S_u}$.
\end{proof}

Since the above response relevance condition applies for all responses
$r = f_D^\cM(q)$ on all queries $q$ by all users $u$, we say that
\multilora, \multiloramerge, and \multiloraunion\ correctly enforce
parameter separation and hence correctly enforce access control for
all users $u$.

%% file: MIAs_app.tex
\section{MIAs against LLMs}\label{app:MIAs_related}
In Section~\ref{sec:auditing}, we defined the Domain Distinguishability Index (DDI) as the average success rate of an adversary playing the Domain Distinguishability game over all domain set pairs. That game is implemented with \emph{membership inference attacks} (MIAs)~\citep{yeom2018privacy, carlini2021extracting, mozaffari2024semantic, shi2023detecting, zhang2024min}: the auditor compares a \emph{member} set drawn from the active domain’s training data with a \emph{non-member} set drawn from some other domain, and tries to tell them apart. The better this separation, the larger the DDI. Here, in this section, we expand on the MIA toolbox that underpins DDI—detailing evaluation metrics and the specific attacks we deploy against LLMs.
More generally, an MIA for an LLM $f$ assigns a
\emph{membership score} $A(x,f)$ to a candidate text $x$.  Thresholding
this score at $\varepsilon$ declares $x$ a member (if $A(x,f)\!\ge\!\varepsilon$)
or a non-member (if $A(x,f)\!<\!\varepsilon$). 

\subsection{Metrics} \label{app:MIA_metrics}
We employ two complementary metrics to quantify the success of our membership inference attacks, as used by prior MIA works~\citep{jayaraman2021revisiting, carlini2022membership, mireshghallah22a}:

\paragraph{(1) Attack ROC curves:}
The Receiver Operating Characteristic (ROC) curve illustrates the trade-off between the True Positive Rate (TPR) and the False Positive Rate (FPR) for the attacks. The FPR measures the proportion of non-member samples that are incorrectly classified as members, while the TPR represents the proportion of member samples that are correctly identified as members. We report the Area Under the ROC Curve (AUC-ROC) as an aggregate metric to assess the overall success of the attacks.  AUC-ROC is a threshold-independent metric, and it shows the probability that a positive instance (member) has higher score than a negative instance (non-member). 

\paragraph{(2) Attack TPR at low FPR:}
This metric is crucial for determining the effectiveness of an attack at confidently identifying members of the training dataset without falsely classifying non-members as members. We focus on low FPR thresholds, specifically 1\%, and 5\%. For instance, the TPR at an FPR of 1\% is calculated by setting the detection threshold so that only 1\% of non-member samples are predicted as members.

\subsection{Existing MIAs}\label{app:existing_MIAs}

\paragraph{LOSS~\citep{yeom2018privacy}:}
The LOSS method utilizes the loss value of model $f(.)$ for the given text $x$ as the membership score; a lower loss suggests that the text was seen during training, so $A(x,f)=\ell(f, x)$.

\paragraph{Ref~\citep{carlini2021extracting}:}
Calculating membership scores based solely on loss values often results in high false negative rates. To improve this, a difficulty calibration method can be employed to account for the intrinsic complexity of $x$. For example, repetitive or common phrases typically yield low loss values. One method of calibrating this input complexity is by using another LLM, $Ref(.)$, assumed to be trained on a similar data distribution. The membership score is then defined as the difference in loss values between the target and reference models, $A(x,f)=\ell(x, f) - \ell(x, Ref)$. In our evaluations, we used the base models (i.e., Llama-3.1-8B and Mistral-0.1-7B) before any fine-tuning as the reference models.

\paragraph{Zlib~\citep{carlini2021extracting}:}
Another method to calibrate the difficulty of a sample is by using its zlib compression size, where more complex sentences have higher compression sizes. The membership score is then calculated by normalizing the loss value by the zlib compression size, $A(x,f)= \frac{\ell(x,f)}{\textit{zlib}(x)}$.

\paragraph{Min-K~\citep{shi2023detecting}:}
This attack hypothesizes that non-member samples often have more tokens assigned lower likelihoods. It first calculates the likelihood of each token as $\text{Min-K\%}_{\text{token}}(x_t) = \log p(x_t|x_{<t})$, for each token $x_t$ given the prefix $x_{<t}$. The membership score is then calculated by averaging over the lowest $K\%$ of tokens with lower likelihood, $A(x, f) = \frac{1}{|\text{min-k\%}|} \sum_{x_i \in min-k\%} \text{Min-K\%}_{\text{token}}(x_t)$.

\paragraph{Min-K++~\citep{zhang2024min}:}
This method improves on Min-K by utilizing the insight that maximum likelihood training optimizes the Hessian trace of likelihood over the training data. It calculates a normalized score for each token $x_t$ given the prefix $x_{<t}$ as $\text{Min-K\%++}_{\text{token}}(x_t) = \frac{\log p(x_t|x_{<t})-\mu_{x_{<t}}}{\sigma_{x_{<t}}}$, where $\mu_{x_{<t}}$ is the mean log probability of the next token across the vocabulary, and $\sigma_{x_{<t}}$ is the standard deviation. The membership score is then aggregated by averaging the scores of the lowest $K\%$ tokens, $A(x, f) = \frac{1}{|\text{min-k\%++}|} \sum_{x_i \in min-k\%} \text{Min-K\%++}_{\text{token}}(x_t)$.

%% file: main.bbl
\begin{thebibliography}{51}
\providecommand{\natexlab}[1]{#1}
\providecommand{\url}[1]{\texttt{#1}}
\expandafter\ifx\csname urlstyle\endcsname\relax
  \providecommand{\doi}[1]{doi: #1}\else
  \providecommand{\doi}{doi: \begingroup \urlstyle{rm}\Url}\fi

\bibitem[Arriaga et~al.(2016)Arriaga, Barbosa, and Farshim]{arriaga16}
Afonso Arriaga, Manuel Barbosa, and Pooya Farshim.
\newblock Private functional encryption: Indistinguishability-based definitions and constructions from obfuscation.
\newblock Cryptology {ePrint} Archive, Paper 2016/018, 2016.
\newblock URL \url{https://eprint.iacr.org/2016/018}.

\bibitem[Bagdasarian et~al.(2024)Bagdasarian, Yi, Ghalebikesabi, Kairouz, Gruteser, Oh, Balle, and Ramage]{bagdasarian24}
Eugene Bagdasarian, Ren Yi, Sahra Ghalebikesabi, Peter Kairouz, Marco Gruteser, Sewoong Oh, Borja Balle, and Daniel Ramage.
\newblock Airgapagent: Protecting privacy-conscious conversational agents, 2024.
\newblock URL \url{https://arxiv.org/abs/2405.05175}.

\bibitem[Bohli and Pashalidis(2011)]{bohli20}
Jens-Matthias Bohli and Andreas Pashalidis.
\newblock Relations among privacy notions.
\newblock \emph{ACM Trans. Inf. Syst. Secur.}, 14\penalty0 (1), 2011.
\newblock URL \url{https://doi.org/10.1145/1952982.1952986}.

\bibitem[B\"{u}ttcher and Clarke(2005)]{buttcher05}
Stefan B\"{u}ttcher and Charles L.~A. Clarke.
\newblock A security model for full-text file system search in multi-user environments.
\newblock In \emph{Proceedings of the 4th Conference on USENIX Conference on File and Storage Technologies - Volume 4}, page~13, USA, 2005. USENIX Association.

\bibitem[Carlini et~al.(2021)Carlini, Tramer, Wallace, Jagielski, Herbert-Voss, Lee, Roberts, Brown, Song, Erlingsson, et~al.]{carlini2021extracting}
Nicholas Carlini, Florian Tramer, Eric Wallace, Matthew Jagielski, Ariel Herbert-Voss, Katherine Lee, Adam Roberts, Tom Brown, Dawn Song, Ulfar Erlingsson, et~al.
\newblock Extracting training data from large language models.
\newblock In \emph{30th USENIX Security Symposium (USENIX Security 21)}, pages 2633--2650, 2021.

\bibitem[Carlini et~al.(2022)Carlini, Chien, Nasr, Song, Terzis, and Tramer]{carlini2022membership}
Nicholas Carlini, Steve Chien, Milad Nasr, Shuang Song, Andreas Terzis, and Florian Tramer.
\newblock Membership inference attacks from first principles.
\newblock In \emph{2022 IEEE symposium on security and privacy (SP)}, pages 1897--1914. IEEE, 2022.

\bibitem[Chan(2025)]{chan25}
Shih-Han Chan.
\newblock Encrypted prompt: Securing llm applications against unauthorized actions, 2025.
\newblock URL \url{https://arxiv.org/abs/2503.23250}.

\bibitem[Chen et~al.(2023)Chen, Mendes, Das, Xu, and Ritter]{chen23a}
Yang Chen, Ethan Mendes, Sauvik Das, Wei Xu, and Alan Ritter.
\newblock Can language models be instructed to protect personal information?, 2023.
\newblock URL \url{https://arxiv.org/abs/2310.02224}.

\bibitem[Dwork et~al.(2006)Dwork, McSherry, Nissim, and Smith]{dwork06}
Cynthia Dwork, Frank McSherry, Kobbi Nissim, and Adam Smith.
\newblock Calibrating noise to sensitivity in private data analysis.
\newblock In \emph{Proceedings of the Third Conference on Theory of Cryptography}, TCC'06, pages 265--284, 2006.

\bibitem[Ferraiolo et~al.(1999)Ferraiolo, Barkley, and Kuhn]{ferraiolo99}
David~F. Ferraiolo, John~F. Barkley, and D.~Richard Kuhn.
\newblock A role-based access control model and reference implementation within a corporate intranet.
\newblock \emph{ACM Trans. Inf. Syst. Secur.}, 2\penalty0 (1):\penalty0 34--64, 1999.
\newblock ISSN 1094-9224.
\newblock URL \url{https://doi.org/10.1145/300830.300834}.

\bibitem[Ferraiolo et~al.(2007)Ferraiolo, Kuhn, and Chandramouli]{ferraiolo07}
David~F. Ferraiolo, D.~Richard Kuhn, and Ramaswamy Chandramouli.
\newblock Role-based access control.
\newblock \emph{Information Security and Privacy Series}, 2007.

\bibitem[Fleshman et~al.(2025)Fleshman, Khan, Marone, and Durme]{fleshman2025adapterswap}
William Fleshman, Aleem Khan, Marc Marone, and Benjamin~Van Durme.
\newblock Adapterswap: Continuous training of llms with data removal and access-control guarantees, 2025.
\newblock URL \url{https://arxiv.org/abs/2404.08417}.

\bibitem[Goldwasser et~al.(1988)Goldwasser, Micali, and Rivest]{goldwasser88}
Shafi Goldwasser, Silvio Micali, and Ronald~L. Rivest.
\newblock A digital signature scheme secure against adaptive chosen-message attacks.
\newblock \emph{SIAM Journal on Computing}, 17\penalty0 (2):\penalty0 281--308, 1988.

\bibitem[Goyal(2023)]{goyal23}
Pawam Goyal.
\newblock Private information retrieval with access control, 2023.

\bibitem[Grattafiori et~al.(2024)Grattafiori, Dubey, Jauhri, Pandey, Kadian, Al-Dahle, Letman, Mathur, Schelten, Vaughan, et~al.]{llama3}
Aaron Grattafiori, Abhimanyu Dubey, Abhinav Jauhri, Abhinav Pandey, Abhishek Kadian, Ahmad Al-Dahle, Aiesha Letman, Akhil Mathur, Alan Schelten, Alex Vaughan, et~al.
\newblock The llama 3 herd of models, 2024.
\newblock URL \url{https://arxiv.org/abs/2407.21783}.

\bibitem[Houlsby et~al.(2019)Houlsby, Giurgiu, Jastrzebski, Morrone, de~Laroussilhe, Gesmundo, Attariyan, and Gelly]{houlsby19}
Neil Houlsby, Andrei Giurgiu, Stanislaw Jastrzebski, Bruna Morrone, Quentin de~Laroussilhe, Andrea Gesmundo, Mona Attariyan, and Sylvain Gelly.
\newblock Parameter-efficient transfer learning for nlp, 2019.
\newblock URL \url{https://arxiv.org/abs/1902.00751}.

\bibitem[Hu et~al.(2021)Hu, Shen, Wallis, Allen-Zhu, Li, Wang, Wang, and Chen]{hu21}
Edward~J. Hu, Yelong Shen, Phillip Wallis, Zeyuan Allen-Zhu, Yuanzhi Li, Shean Wang, Lu~Wang, and Weizhu Chen.
\newblock Lora: Low-rank adaptation of large language models, 2021.
\newblock URL \url{https://arxiv.org/abs/2106.09685}.

\bibitem[Jayaraman et~al.(2021)Jayaraman, Wang, Knipmeyer, Gu, and Evans]{jayaraman2021revisiting}
Bargav Jayaraman, Lingxiao Wang, Katherine Knipmeyer, Quanquan Gu, and David Evans.
\newblock Revisiting membership inference under realistic assumptions.
\newblock \emph{Proceedings on Privacy Enhancing Technologies}, 2021.

\bibitem[Jiang et~al.(2023)Jiang, Sablayrolles, Mensch, Bamford, Chaplot, de~las Casas, Bressand, Lengyel, Lample, Saulnier, Lavaud, Lachaux, Stock, Scao, Lavril, Wang, Lacroix, and Sayed]{mistral}
Albert~Q. Jiang, Alexandre Sablayrolles, Arthur Mensch, Chris Bamford, Devendra~Singh Chaplot, Diego de~las Casas, Florian Bressand, Gianna Lengyel, Guillaume Lample, Lucile Saulnier, Lélio~Renard Lavaud, Marie-Anne Lachaux, Pierre Stock, Teven~Le Scao, Thibaut Lavril, Thomas Wang, Timothée Lacroix, and William~El Sayed.
\newblock Mistral 7b, 2023.
\newblock URL \url{https://arxiv.org/abs/2310.06825}.

\bibitem[Jin et~al.(2019)Jin, Dhingra, Liu, Cohen, and Lu]{jin-etal-2019-pubmedqa}
Qiao Jin, Bhuwan Dhingra, Zhengping Liu, William Cohen, and Xinghua Lu.
\newblock {P}ub{M}ed{QA}: A dataset for biomedical research question answering.
\newblock In \emph{Proceedings of the 2019 Conference on Empirical Methods in Natural Language Processing and the 9th International Joint Conference on Natural Language Processing (EMNLP-IJCNLP)}, pages 2567--2577. Association for Computational Linguistics, 2019.
\newblock \doi{10.18653/v1/D19-1259}.
\newblock URL \url{https://aclanthology.org/D19-1259/}.

\bibitem[Kaddour et~al.(2023)Kaddour, Harris, Mozes, Bradley, Raileanu, and McHardy]{kaddour23}
Jean Kaddour, Joshua Harris, Maximilian Mozes, Herbie Bradley, Roberta Raileanu, and Robert McHardy.
\newblock Challenges and applications of large language models, 2023.
\newblock URL \url{https://arxiv.org/abs/2307.10169}.

\bibitem[Lewis et~al.(2004)Lewis, Yang, Rose, and Li]{rcv1}
David~D Lewis, Yiming Yang, Tony~G Rose, and Fan Li.
\newblock Rcv1: A new benchmark collection for text categorization research.
\newblock \emph{Journal of machine learning research}, 5:\penalty0 361--397, 2004.

\bibitem[Lewis et~al.(2020)Lewis, Perez, Piktus, Petroni, Karpukhin, Goyal, K\"{u}ttler, Lewis, Yih, Rockt\"{a}schel, Riedel, and Kiela]{lewis20}
Patrick Lewis, Ethan Perez, Aleksandra Piktus, Fabio Petroni, Vladimir Karpukhin, Naman Goyal, Heinrich K\"{u}ttler, Mike Lewis, Wen-tau Yih, Tim Rockt\"{a}schel, Sebastian Riedel, and Douwe Kiela.
\newblock Retrieval-augmented generation for knowledge-intensive nlp tasks.
\newblock In \emph{Proceedings of the 34th International Conference on Neural Information Processing Systems}, NIPS '20, Red Hook, NY, USA, 2020. Curran Associates Inc.
\newblock ISBN 9781713829546.

\bibitem[Li et~al.(2024)Li, Pan, Gopal, Yue, Berrios, Gatti, Li, Dombrowski, Goel, Phan, et~al.]{wmdp}
Nathaniel Li, Alexander Pan, Anjali Gopal, Summer Yue, Daniel Berrios, Alice Gatti, Justin~D Li, Ann-Kathrin Dombrowski, Shashwat Goel, Long Phan, et~al.
\newblock The {WMDP} benchmark: Measuring and reducing malicious use with unlearning.
\newblock In \emph{Proceedings of the 41st International Conference on Machine Learning}, volume 235 of \emph{Proceedings of Machine Learning Research}, pages 28525--28550. PMLR, 2024.
\newblock URL \url{https://proceedings.mlr.press/v235/li24bc.html}.

\bibitem[Liu et~al.(2025)Liu, Wang, Xiao, and Chen]{liu25}
Qin Liu, Fei Wang, Chaowei Xiao, and Muhao Chen.
\newblock Sudolm: Learning access control of parametric knowledge with authorization alignment, 2025.
\newblock URL \url{https://arxiv.org/abs/2410.14676}.

\bibitem[Liu et~al.(2024)Liu, Deng, Xu, Li, Zheng, Zhang, Zhao, Zhang, Wang, and Liu]{liu2024jailbreaking}
Yi~Liu, Gelei Deng, Zhengzi Xu, Yuekang Li, Yaowen Zheng, Ying Zhang, Lida Zhao, Tianwei Zhang, Kailong Wang, and Yang Liu.
\newblock Jailbreaking chatgpt via prompt engineering: An empirical study, 2024.
\newblock URL \url{https://arxiv.org/abs/2305.13860}.

\bibitem[Mehrotra et~al.(2024)Mehrotra, Zampetakis, Kassianik, Nelson, Anderson, Singer, and Karbasi]{NEURIPS2024_70702e8c}
Anay Mehrotra, Manolis Zampetakis, Paul Kassianik, Blaine Nelson, Hyrum Anderson, Yaron Singer, and Amin Karbasi.
\newblock Tree of attacks: Jailbreaking black-box llms automatically.
\newblock In A.~Globerson, L.~Mackey, D.~Belgrave, A.~Fan, U.~Paquet, J.~Tomczak, and C.~Zhang, editors, \emph{Advances in Neural Information Processing Systems}, volume~37, pages 61065--61105. Curran Associates, Inc., 2024.
\newblock URL \url{https://proceedings.neurips.cc/paper_files/paper/2024/file/70702e8cbb4890b4a467b984ae59828a-Paper-Conference.pdf}.

\bibitem[Mireshghallah et~al.(2022)Mireshghallah, Goyal, Uniyal, Berg{-}Kirkpatrick, and Shokri]{mireshghallah22a}
Fatemehsadat Mireshghallah, Kartik Goyal, Archit Uniyal, Taylor Berg{-}Kirkpatrick, and Reza Shokri.
\newblock Quantifying privacy risks of masked language models using membership inference attacks.
\newblock In \emph{Proceedings of the 2022 Conference on Empirical Methods in Natural Language Processing}, pages 8332--8347, 2022.

\bibitem[Mozaffari and Marathe(2024)]{mozaffari2024semantic}
Hamid Mozaffari and Virendra~J Marathe.
\newblock Semantic membership inference attack against large language models.
\newblock \emph{arXiv preprint arXiv:2406.10218}, 2024.

\bibitem[{NIST SP Joint Task Force}(2020)]{nistsp20}
{NIST SP Joint Task Force}.
\newblock Security and privacy controls for information systems and organizations, {\it nist sp 800-53 rev. 5}, 2020.
\newblock URL \url{https://csrc.nist.gov/pubs/sp/800/53/r5/upd1/final}.

\bibitem[Ostapenko et~al.(2024)Ostapenko, Su, Ponti, Charlin, Roux, Pereira, Caccia, and Sordoni]{ostapenko2024modularllmsbuildingreusing}
Oleksiy Ostapenko, Zhan Su, Edoardo~Maria Ponti, Laurent Charlin, Nicolas~Le Roux, Matheus Pereira, Lucas Caccia, and Alessandro Sordoni.
\newblock Towards modular llms by building and reusing a library of loras, 2024.
\newblock URL \url{https://arxiv.org/abs/2405.11157}.

\bibitem[{OWASP GenAI Security Project}(2025)]{owasp25}
{OWASP GenAI Security Project}.
\newblock Llm08:2025 vector and embedding weaknesses, 2025.
\newblock URL \url{https://genai.owasp.org/llmrisk/llm082025-vector-and-embedding-weaknesses/#:~:text=1.\%20Permission\%20and\%20access\%20control}.

\bibitem[Rein et~al.(2023)Rein, Hou, Stickland, Petty, Pang, Dirani, Michael, and Bowman]{gpqa}
David Rein, Betty~Li Hou, Asa~Cooper Stickland, Jackson Petty, Richard~Yuanzhe Pang, Julien Dirani, Julian Michael, and Samuel~R. Bowman.
\newblock Gpqa: A graduate-level google-proof q\&a benchmark, 2023.
\newblock URL \url{https://arxiv.org/abs/2311.12022}.

\bibitem[Shang et~al.(2024)Shang, Yao, Yao, Su, Fan, Zhang, and Jiang]{ShangYYSFZJ24}
Shang Shang, Zhongjiang Yao, Yepeng Yao, Liya Su, Zijing Fan, Xiaodan Zhang, and Zhengwei Jiang.
\newblock Intentobfuscator: A jailbreaking method via confusing llm with prompts.
\newblock In \emph{ESORICS}, pages 146--165, 2024.
\newblock URL \url{https://doi.org/10.1007/978-3-031-70903-6_8}.

\bibitem[Shen et~al.(2025)Shen, Qiu, Kurmanji, Iacob, Sani, Chen, Cancedda, and Lane]{shen25}
William~F. Shen, Xinchi Qiu, Meghdad Kurmanji, Alex Iacob, Lorenzo Sani, Yihong Chen, Nicola Cancedda, and Nicholas~D. Lane.
\newblock Lunar: Llm unlearning via neural activation redirection, 2025.
\newblock URL \url{https://arxiv.org/abs/2502.07218}.

\bibitem[Shi et~al.(2023)Shi, Ajith, Xia, Huang, Liu, Blevins, Chen, and Zettlemoyer]{shi2023detecting}
Weijia Shi, Anirudh Ajith, Mengzhou Xia, Yangsibo Huang, Daogao Liu, Terra Blevins, Danqi Chen, and Luke Zettlemoyer.
\newblock Detecting pretraining data from large language models.
\newblock \emph{arXiv preprint arXiv:2310.16789}, 2023.

\bibitem[Shokri et~al.(2017)Shokri, Stronati, Song, and Shmatikov]{shokri2017membership}
Reza Shokri, Marco Stronati, Congzheng Song, and Vitaly Shmatikov.
\newblock Membership inference attacks against machine learning models.
\newblock In \emph{2017 IEEE symposium on security and privacy (SP)}, pages 3--18. IEEE, 2017.

\bibitem[Stoica et~al.(2024)Stoica, Ramesh, Ecsedi, Choshen, and Hoffman]{stoica24}
George Stoica, Pratik Ramesh, Boglarka Ecsedi, Leshem Choshen, and Judy Hoffman.
\newblock Model merging with {SVD} to tie the {Knots}, 2024.
\newblock URL \url{https://arxiv.org/abs/2410.19735}.

\bibitem[Vu et~al.(2022)Vu, Lester, Constant, Al-Rfou, and Cer]{vu2022spotbetterfrozenmodel}
Tu~Vu, Brian Lester, Noah Constant, Rami Al-Rfou, and Daniel Cer.
\newblock Spot: Better frozen model adaptation through soft prompt transfer, 2022.
\newblock URL \url{https://arxiv.org/abs/2110.07904}.

\bibitem[Wei et~al.(2023)Wei, Haghtalab, and Steinhardt]{NEURIPS2023_fd661313}
Alexander Wei, Nika Haghtalab, and Jacob Steinhardt.
\newblock Jailbroken: How does llm safety training fail?
\newblock In A.~Oh, T.~Naumann, A.~Globerson, K.~Saenko, M.~Hardt, and S.~Levine, editors, \emph{Advances in Neural Information Processing Systems}, volume~36, pages 80079--80110. Curran Associates, Inc., 2023.
\newblock URL \url{https://proceedings.neurips.cc/paper_files/paper/2023/file/fd6613131889a4b656206c50a8bd7790-Paper-Conference.pdf}.

\bibitem[Wei et~al.(2024)Wei, Karina, Chung, Jiao, Papay, Glaese, Schulman, and Fedus]{simpleqa}
Jason Wei, Nguyen Karina, Hyung~Won Chung, Yunxin~Joy Jiao, Spencer Papay, Amelia Glaese, John Schulman, and William Fedus.
\newblock Measuring short-form factuality in large language models, 2024.
\newblock URL \url{https://arxiv.org/abs/2411.04368}.

\bibitem[Xu et~al.(2023)Xu, Xie, Qin, Tao, and Wang]{xu23}
Lingling Xu, Haoran Xie, Si-Zhao~Joe Qin, Xiaohui Tao, and Fu~Lee Wang.
\newblock Parameter-efficient fine-tuning methods for pretrained language models: A critical review and assessment, 2023.
\newblock URL \url{https://arxiv.org/abs/2312.12148}.

\bibitem[Yadav et~al.(2023)Yadav, Tam, Choshen, Raffel, and Bansal]{yadav23}
Prateek Yadav, Derek Tam, Leshem Choshen, Colin Raffel, and Mohit Bansal.
\newblock Ties-merging: Resolving interference when merging models, 2023.
\newblock URL \url{https://arxiv.org/abs/2306.01708}.

\bibitem[Yeom et~al.(2018)Yeom, Giacomelli, Fredrikson, and Jha]{yeom2018privacy}
Samuel Yeom, Irene Giacomelli, Matt Fredrikson, and Somesh Jha.
\newblock Privacy risk in machine learning: Analyzing the connection to overfitting, 2018.

\bibitem[Yin et~al.(2024)Yin, Ye, and Durrett]{yin24}
Fangcong Yin, Xi~Ye, and Greg Durrett.
\newblock Lofit: Localized fine-tuning on {LLM} representations.
\newblock In \emph{Advances in Neural Information Processing Systems 38: Annual Conference on Neural Information Processing Systems}, 2024.

\bibitem[Yu et~al.(2024)Yu, Yu, Yu, Huang, and Li]{yu24}
Le~Yu, Bowen Yu, Haiyang Yu, Fei Huang, and Yongbin Li.
\newblock Language models are super mario: Absorbing abilities from homologous models as a free lunch, 2024.
\newblock URL \url{https://arxiv.org/abs/2311.03099}.

\bibitem[Zhang et~al.(2024)Zhang, Sun, Yeats, Ouyang, Kuo, Zhang, Yang, and Li]{zhang2024min}
Jingyang Zhang, Jingwei Sun, Eric Yeats, Yang Ouyang, Martin Kuo, Jianyi Zhang, Hao Yang, and Hai Li.
\newblock Min-k\%++: Improved baseline for detecting pre-training data from large language models.
\newblock \emph{arXiv preprint arXiv:2404.02936}, 2024.

\bibitem[Zhao et~al.(2025)Zhao, Zhou, Li, Tang, Wang, Hou, Min, Zhang, Zhang, Dong, Du, Yang, Chen, Chen, Jiang, Ren, Li, Tang, Liu, Liu, Nie, and Wen]{zhao25}
Wayne~Xin Zhao, Kun Zhou, Junyi Li, Tianyi Tang, Xiaolei Wang, Yupeng Hou, Yingqian Min, Beichen Zhang, Junjie Zhang, Zican Dong, Yifan Du, Chen Yang, Yushuo Chen, Zhipeng Chen, Jinhao Jiang, Ruiyang Ren, Yifan Li, Xinyu Tang, Zikang Liu, Peiyu Liu, Jian-Yun Nie, and Ji-Rong Wen.
\newblock A survey of large language models, 2025.
\newblock URL \url{https://arxiv.org/abs/2303.18223}.

\bibitem[Zhao et~al.(2018)Zhao, Li, Shen, Liang, and Wu]{Zhao_2018_ECCV}
Xiangyun Zhao, Haoxiang Li, Xiaohui Shen, Xiaodan Liang, and Ying Wu.
\newblock A modulation module for multi-task learning with applications in image retrieval.
\newblock In \emph{Proceedings of the European Conference on Computer Vision (ECCV)}, 2018.

\bibitem[Zhao et~al.(2024)Zhao, Shen, Zhu, Li, Su, Wang, Kuang, and Wu]{zhao24}
Ziyu Zhao, Tao Shen, Didi Zhu, Zexi Li, Jing Su, Xuwu Wang, Kun Kuang, and Fei Wu.
\newblock Merging loras like playing lego: Pushing the modularity of lora to extremes through rank-wise clustering, 2024.
\newblock URL \url{https://arxiv.org/abs/2409.16167}.

\bibitem[Zhong et~al.(2025)Zhong, Lentz, Narodytska, Szekeres, and Rong]{zhong2025honeybeeefficientrolebasedaccess}
Hongbin Zhong, Matthew Lentz, Nina Narodytska, Adriana Szekeres, and Kexin Rong.
\newblock Honeybee: Efficient role-based access control for vector databases via dynamic partitioning, 2025.
\newblock URL \url{https://arxiv.org/abs/2505.01538}.

\end{thebibliography}
